%% file: Main_Manuscript.tex
\documentclass[acmsmall]{acmart}

\AtBeginDocument{%
  }

\setcopyright{rightsretained}
\acmYear{2024}
\acmDOI{10.1145/3677074}

\acmJournal{PACMHCI}
\acmVolume{8} 
\acmNumber{CHI PLAY} 
\acmArticle{309}
\acmMonth{10}

\input{sections/99-Commands}

\makeatletter
\gdef\@copyrightpermission{
  \begin{minipage}{0.2\columnwidth}
   \href{https://creativecommons.org/licenses/by-sa/4.0/}{\includegraphics[width=0.90\textwidth]{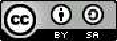}}
  \end{minipage}\hfill
  \begin{minipage}{0.8\columnwidth}
   \href{https://creativecommons.org/licenses/by-sa/4.0/}{This work is licensed under a Creative Commons Attribution-ShareAlike International 4.0 License.}
  \end{minipage}
  \vspace{5pt}
}
\makeatother

\begin{document}

\title[Deceptive Design in a Free-to-Play Game Transition]{From Motivating to Manipulative: The Use of Deceptive Design in a Game's Free-to-Play Transition}

\author{Hilda Hadan}
\email{hhadan@uwaterloo.ca}
\orcid{https://orcid.org/0000-0002-5911-1405}
\affiliation{
    \institution{Stratford School of Interaction Design and Business, University of Waterloo}
    \city{Waterloo}
    \country{Canada}
}

\author{Sabrina Alicia Sgandurra}
\email{sasgandu@uwaterloo.ca}
\orcid{https://orcid.org/0000-0003-3187-263X}
\affiliation{
    \institution{English Language and Literature, Faculty of Arts, University of Waterloo}
    \city{Waterloo}
    \country{Canada}
}

\author{Leah Zhang-Kennedy}
\email{lzhangke@uwaterloo.ca}
\orcid{https://orcid.org/0000-0002-0756-0022}
\affiliation{
    \institution{Stratford School of Interaction Design and Business, University of Waterloo}
    \city{Waterloo}
    \country{Canada}
}

\author{Lennart E. Nacke}
\email{lennart.nacke@acm.org}
\orcid{https://orcid.org/0000-0003-4290-8829}
\affiliation{
    \institution{Stratford School of Interaction Design and Business, University of Waterloo}
    \city{Waterloo}
    \country{Canada}
}


\renewcommand{\shortauthors}{Hilda Hadan, Sabrina Alicia Sgandurra, Leah Zhang-Kennedy, \& Lennart E. Nacke}

\begin{abstract}
\input{sections/00-Abstract}

\end{abstract}




\received{February 2024}
\received[revised]{June 2024}
\received[accepted]{July 2024}

\maketitle

\section{Introduction}
\label{sec:introduction}
\input{sections/01-Introduction}

\section{Related Work}
\label{sec:related-work}
\input{sections/02-RelatedWork}

\section{Methodology}
\label{sec:methodology}
\input{sections/03-Methodology}

\section{Results}
\label{sec:results}
\input{sections/04-Results}

\section{Discussion}
\label{sec:discussion}
\input{sections/05-Discussion}

\section{Conclusion}
\label{sec:conclusion}
\input{sections/06-Conclusion}

\begin{acks}

We thank the anonymous Reddit post creators for sharing their OW2 experiences. We also thank graduate researchers Derrick Wang and Joseph Tu, and Dr. Eugene Kukshinov for their participation and contribution in the collaborative review session, and we thank graduate researcher Michaela Valiquette for her contribution to the thematic analysis. 

The research was supported by the Natural Sciences and Engineering Research Council of Canada (NSERC) Discovery Grant (\#RGPIN-2022-03353 and \#RGPIN-2023-03705), the Social Sciences and Humanities Research Council of Canada (SSHRC) Insight Grant (\#435-2022-0476), the Canada Foundation for Innovation (CFI) JELF Grant (\#41844). Any opinions, findings, and conclusions or recommendations expressed in this material are those of the author(s) and do not necessarily reflect the views of the NSERC, SSHRC, the CFI, nor the University of Waterloo.

\end{acks}

\bibliographystyle{ACM-Reference-Format}
\bibliography{paper}

\appendix
\input{sections/98-Appendix}  

\end{document}

%% file: sections/99-Commands.tex
\usepackage{float}

\usepackage[shortlabels]{enumitem}

\usepackage{xspace} 
\newcommand{\bmt}{business model transition\xspace}

\newcommand{\BP}{Battle Pass\xspace}

\usepackage{microtype}  
\newcommand{\varname}[1]{\texttt{\textls[-40]{#1}}}

\usepackage{multirow}
\usepackage{adjustbox}
\usepackage{fontawesome}

\usepackage{colortbl}

\usepackage{soul}

\usepackage{xcolor,colortbl}
\definecolor{palered}{HTML}{ffadad}
\definecolor{paleorange}{HTML}{ffd6a5}
\definecolor{paleyellow}{HTML}{fdffb6}
\definecolor{palegreen}{HTML}{caffbf}
\definecolor{palecyan}{HTML}{9bf6ff}
\definecolor{paleblue}{HTML}{a0c4ff}
\definecolor{palepurple}{HTML}{bdb2ff}
\definecolor{palepink}{HTML}{ffc6ff}
\definecolor{palewhite}{HTML}{fffffc}

\definecolor{redwood}{HTML}{000000} 

\definecolor{ferngreen}{HTML}{000000}  

\definecolor{skyblue}{HTML}{000000}  
\newcommand{\UImethod}[1]{\textcolor{skyblue}{#1}}
\newcommand{\Postmethod}[1]{\textcolor{skyblue}{#1}}
\definecolor{amethyst}{HTML}{000000}  
\newcommand{\discussion}[1]{\textcolor{amethyst}{#1}}
\definecolor{UTorange}{HTML}{000000}  
\newcommand{\minor}[1]{\textcolor{UTorange}{#1}}

\newenvironment{quoting}[1][]
  {\par\small
   \begin{list}{}{%
     \setlength{\topsep}{1em}
     \setlength{\partopsep}{0pt}
     \setlength{\itemsep}{0pt}
     \setlength{\parsep}{0pt}
     \leftmargin=1cm \rightmargin=1cm}
   \item\relax
   \begingroup\itshape
   \ifx\relax#1\relax\else\textbf{#1:}\ \fi}
  {\par\endgroup\end{list}\addvspace{1em}}

%% file: sections/00-Abstract.tex
Over the last decade, the free-to-play (F2P) game business model has gained popularity in the games industry. We examine the role of deceptive design during a game's transition to F2P and its impacts on players. Our analysis focuses on game mechanics and a Reddit analysis of the Overwatch (OW) series after it transitioned to an F2P model. Our study identifies nine game mechanics that use deceptive design patterns. We also identify factors contributing to a negative gameplay experience. Business model transitions in games present possibilities for problematic practices. Our findings identify the need for game developers and publishers to balance player investments and fairness of rewards. A game's successful transition depends on maintaining fundamental components of player motivation and ensuring transparent communication. Compared to existing taxonomies in other media, games need a comprehensive classification of deceptive design. We emphasize the importance of understanding player perceptions and the impact of deceptive practices in future research.

%% file: sections/01-Introduction.tex

In the past decade, the Free-to-Play (F2P) business model has become popular in the video game industry~\cite{statistics2022gaming}. Several games companies have considered transitioning their games from the Buy-to-Play (B2P) or the Pay-to-Play (P2P) models to the F2P model~\cite{rizani2020analysis, newham2022consequences,evans2016economics} because it can lower player entry barriers (i.e., game purchase cost in B2P; subscription fees in P2P), and offer significant profitability for developers and publishers~\cite{macinnes2002business, massarczyk2019economic, sanchez2022welfare}. F2P games allow players to join for free, but rely on \textit{micro-transactions} as the main revenue source~\cite{vanhatupa2011business, massarczyk2019economic, petrovskaya2021predatory}. This need to use microtransactions as a revenue source opens the F2P business model up for potential abuse~\cite{petrovskaya2021predatory}. Beyond microtransactions, many game mechanics that incentivize player engagement can also be detrimental to players' economical, social and psychological well-being when deceptive design (also known as ``dark patterns'')\footnote{We follow the ACM Diversity and Inclusion Council's guideline for inclusive language and adopt the term ``deceptive design'' instead of ``dark patterns'' in our study. See: \url{https://www.acm.org/diversity-inclusion/words-matter}} is implemented~\cite{zagal2013dark,king20233d,petrovskaya2021predatory}. For example, Epic Games was penalized by the US Federal Trade Commission (FTC) in 2022 for using ``dark patterns'' to drive unwanted purchases in the popular game Fortnite~\cite{FTC2022Epics}. Although previous studies have identified and classified deceptive design in games~\cite{zagal2013dark, lewis2014irresistible, aagaard2022game}, the use of deceptive design practices and their effects on players during games' business model transitions has received little attention.

\textit{Deceptive designs}~\cite{zagal2013dark,Brignull2023book} are user interface design practices that trick users into doing things to benefit businesses~\cite{mathur2021makes, Brignull2010deceptive}. When used in games, such designs can cause negative player experiences, such as distorting their sense of time, swaying how much money they spend playing the game, and influencing their behaviour through psychological and emotional manipulation~\cite{zagal2013dark,petrovskaya2021predatory,king20233d}. Research has found common game mechanics that involve deceptive design patterns, including lootboxes (purchasable random items)~\cite{king20233d,petrovskaya2021predatory,neely2021come}, battle pass (pay-to-skip or grind)~\cite{petrovskaya2021predatory}, and in-game currencies (obscures the real cost of items)~\cite{petrovskaya2021predatory,zagal2013dark,fitton2019F2P}. Beyond specific game mechanics, prior research has explored the viability of B2P and P2P games' transition to F2P from publisher's point of view~\cite{newham2022consequences,rizani2020analysis}. This lens neglected the perceptions and experiences of players. Certain design practices implemented during such transitions often appeared to be unfair and aggressive to players, leading to a degraded player experience~\cite{rizani2020analysis}.

We believe that an in-depth investigation of players' perceptions and experiences on the use of design practices in a game's \bmt is crucial. Player perspective is a valuable source for understanding the impacts of game mechanics~\cite{petrovskaya2021predatory, Shokrizade2013f2p, lin2011cash}. This investigation can shed light on commonly adopted deceptive designs during such transitions and their consequences on players. A richer understanding in this area can contribute to prioritizing ethical design in future game business model transitions, promoting successful and mutually beneficial outcomes for both publishers and players. Therefore, we answer the following Research Questions (RQs): 

\begin{enumerate}[label= \textbf{RQ\arabic*:}]
    \item How do players perceive the role of deceptive game mechanics resulting from the \bmt?
    \item What elements contribute to a satisfying and non-manipulative player experience?
\end{enumerate}

For this research, we study a popular multiplayer first-person shooter (FPS) game series, Overwatch (OW), because it recently transitioned from a B2P to F2P model. While F2P can offer motivating, relaxing, social experiences~\cite{frommel2022daily,marczewski2017ethics}, the developers of OW2 have been heavily criticized for how they handled this transition\footnote{Overwatch 2 was accused of ``deliberately misleading'' players with the newest battle pass:~\url{https://www.rockpapershotgun.com/overwatch-2-accused-of-deliberately-misleading-players-with-the-newest-battle-pass}}. In particular, we chose the OW game series as our research context for two reasons. First, OW's recent \bmt has led to the integration of many deceptive design practices commonly used in F2P games (as outlined in~\autoref{tab:game-mechanics-summary}). Therefore, studying OW will allow us to understand the effect of these design practices on player experience and map these insights onto other, similarly-designed F2P games. Second, OW presents an opportune moment for studying a group of players impacted by a game before and after the integration of deceptive design practices. Overwatch players---especially those that have played both OW1 and OW2 for years---are likely able to compare the changes in design practices that the \bmt triggered. 

\begin{figure}[!t]
    \centering
    \begin{minipage}{0.48\textwidth}
        \centering
        \includegraphics[width=\textwidth]{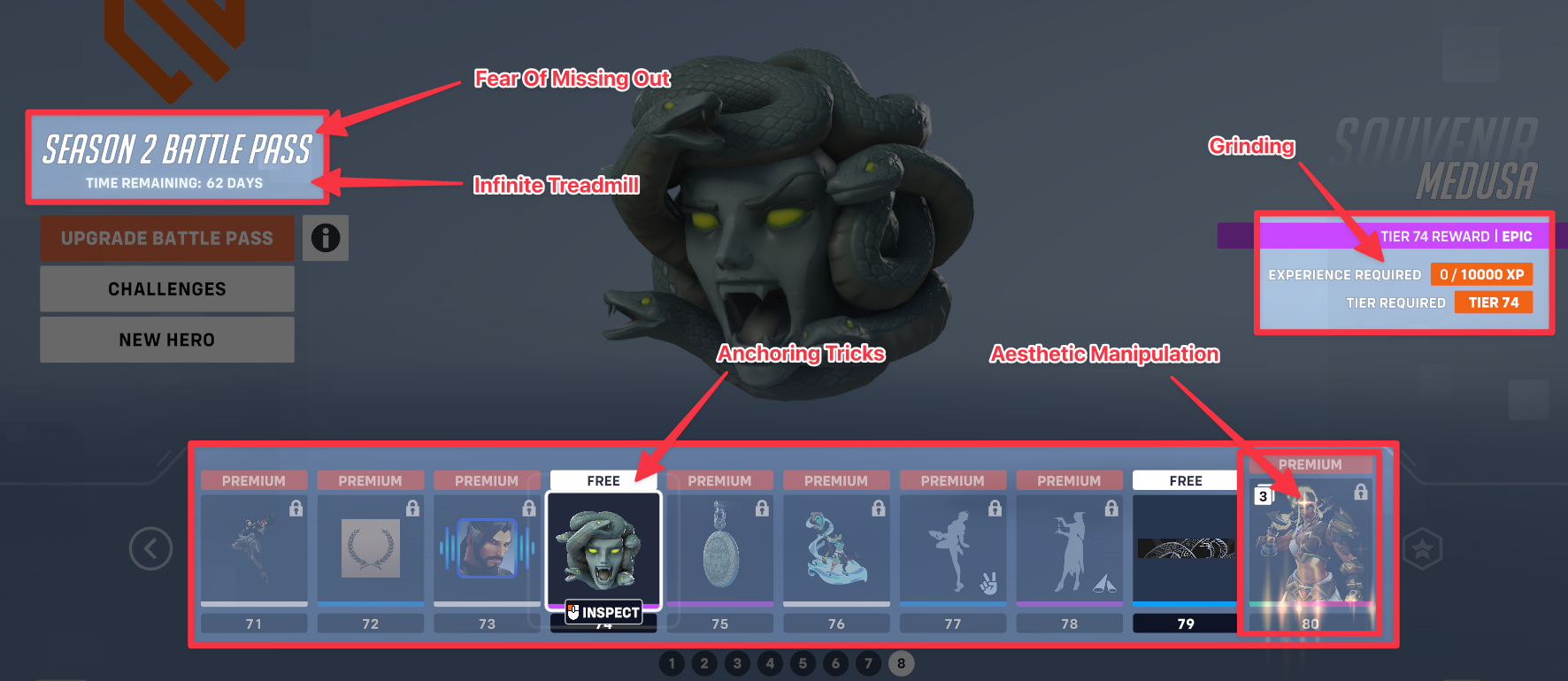}   
        \caption{\BP from OW2 season 2. Each level requires 10,000 XPs (\textit{Grinding}) and rewards players with an item (\textit{Aesthetic Manipulation}). A total of 80 ``premium'' items are only for players who paid premium fee (\textit{Anchoring Tricks}). The progress resets every season~\cite{OW2seasonduration} (\textit{Fear of Missing Out} and \textit{Infinite Treadmill}).
        }
        \Description{\BP from OW2 season 2. Each level requires 10,000 XPs and rewards players with an item. ``Premium'' items are only for premium \BP players. The progress resets every season~\cite{OW2seasonduration}.}
        \label{fig:BattlePass-screenshot}
    \end{minipage}\hfill
    \begin{minipage}{0.48\textwidth}
        \centering
        \includegraphics[width=\textwidth]{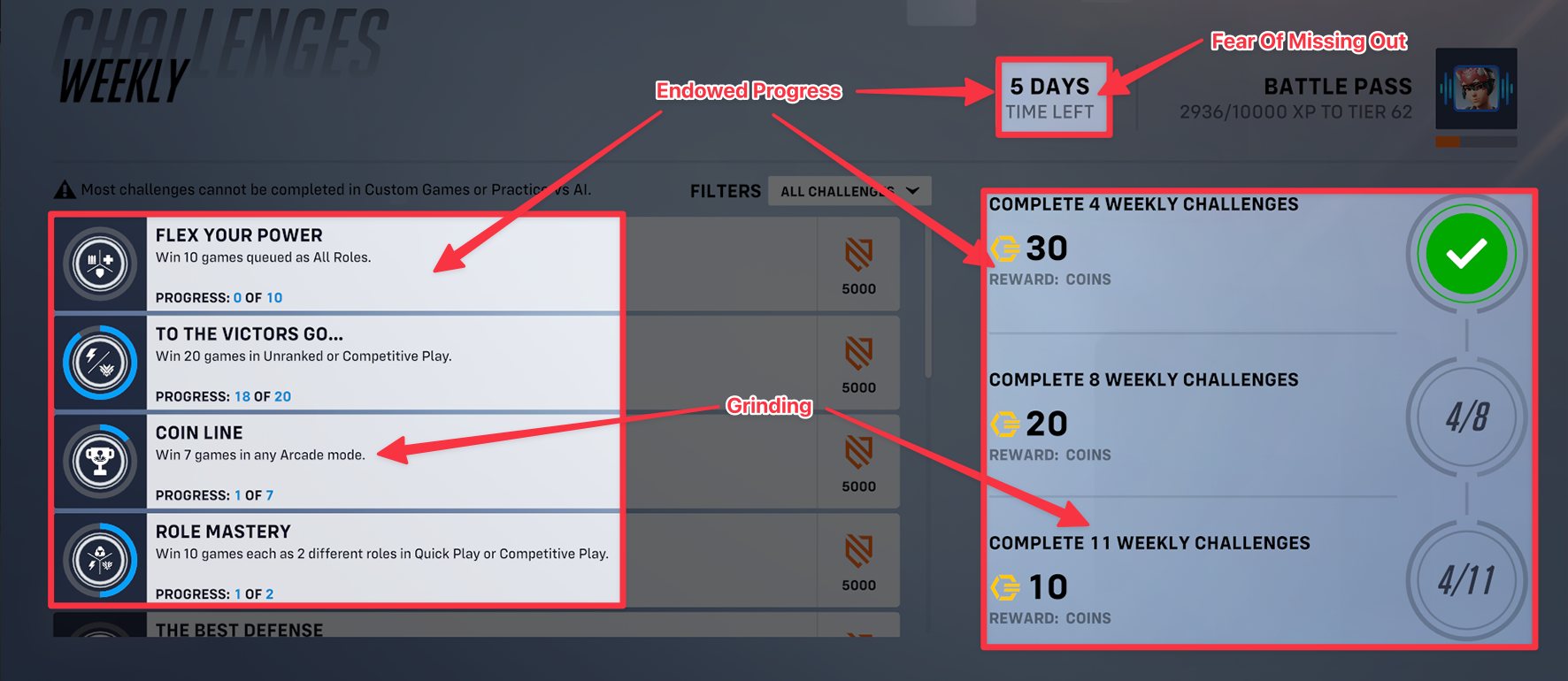}
        \caption{Weekly Challenges from OW2. The completion of each challenge offers 5,000 XPs (\textit{Grinding}). Upon the completion of 4, 8, and 11 challenges, players can obtain 30, 20, and 10 OW coins (\textit{Endowed Progress)}. The progress resets every week (\textit{Fear of Missing Out}).}
        \Description{Weekly Challenges from OW2. The completion of each challenge offers 5,000 XPs. Upon the completion of 4, 8, and 11 challenges, players can obtain 30, 20, and 10 OW coins. The progress resets every week.}
        \label{fig:Challenges-screenshot}
    \end{minipage}
\end{figure}
\begin{figure}[!t]
    \centering
    \begin{minipage}{0.48\textwidth}
        \centering
        \includegraphics[width=\textwidth]{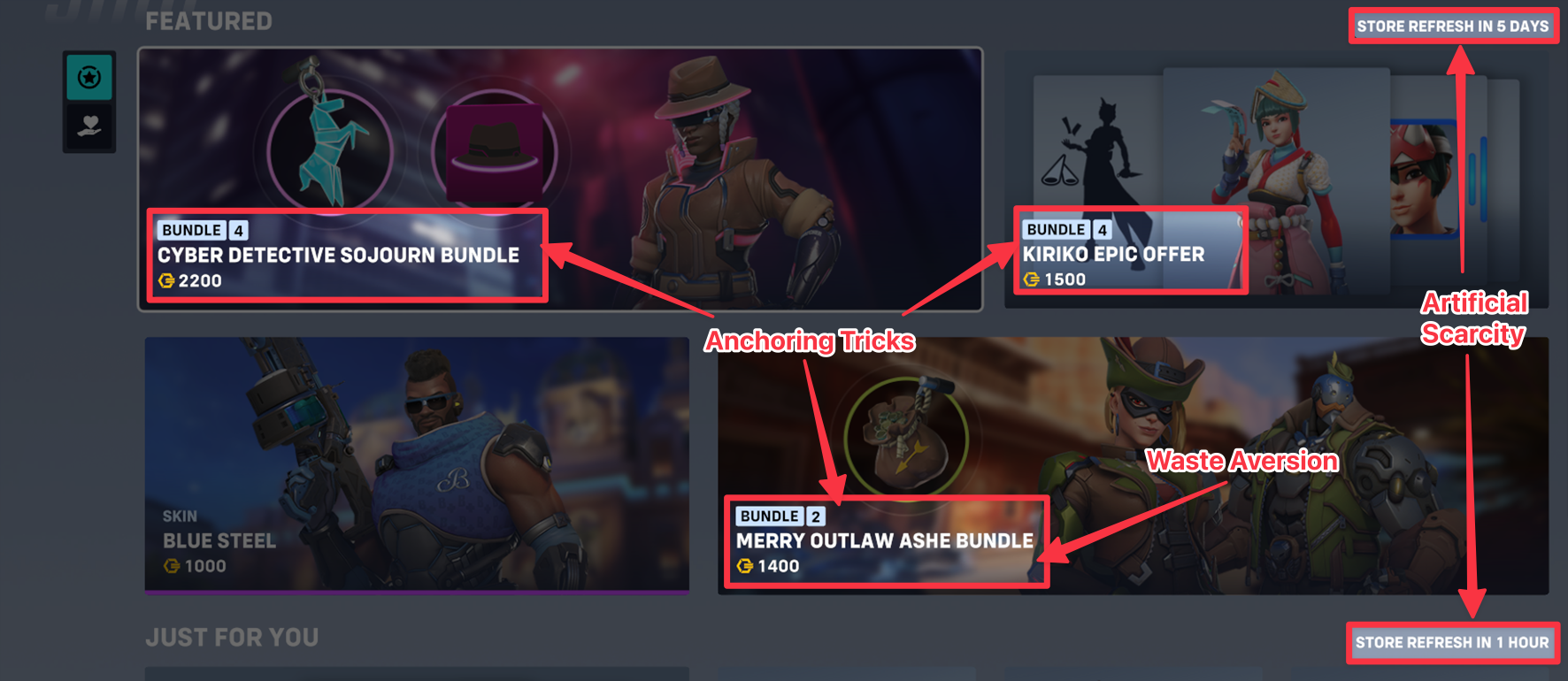}   
        \caption{OW2 Game Shop from season 2 with bundles (\textit{Anchoring Tricks}) and ``featured'' items in a price that is different from amount of purchasable coins (\textit{Waste Aversion}). Items refreshes periodically (\textit{Artificial Scarcity}). 
        }
        \Description{OW2 Game Shop from season 2 with ``featured'' items that refreshes periodically. 
        }
        \label{fig:Shop-screenshot}
    \end{minipage}\hfill
    \begin{minipage}{0.48\textwidth}
        \centering
        \includegraphics[width=\textwidth]{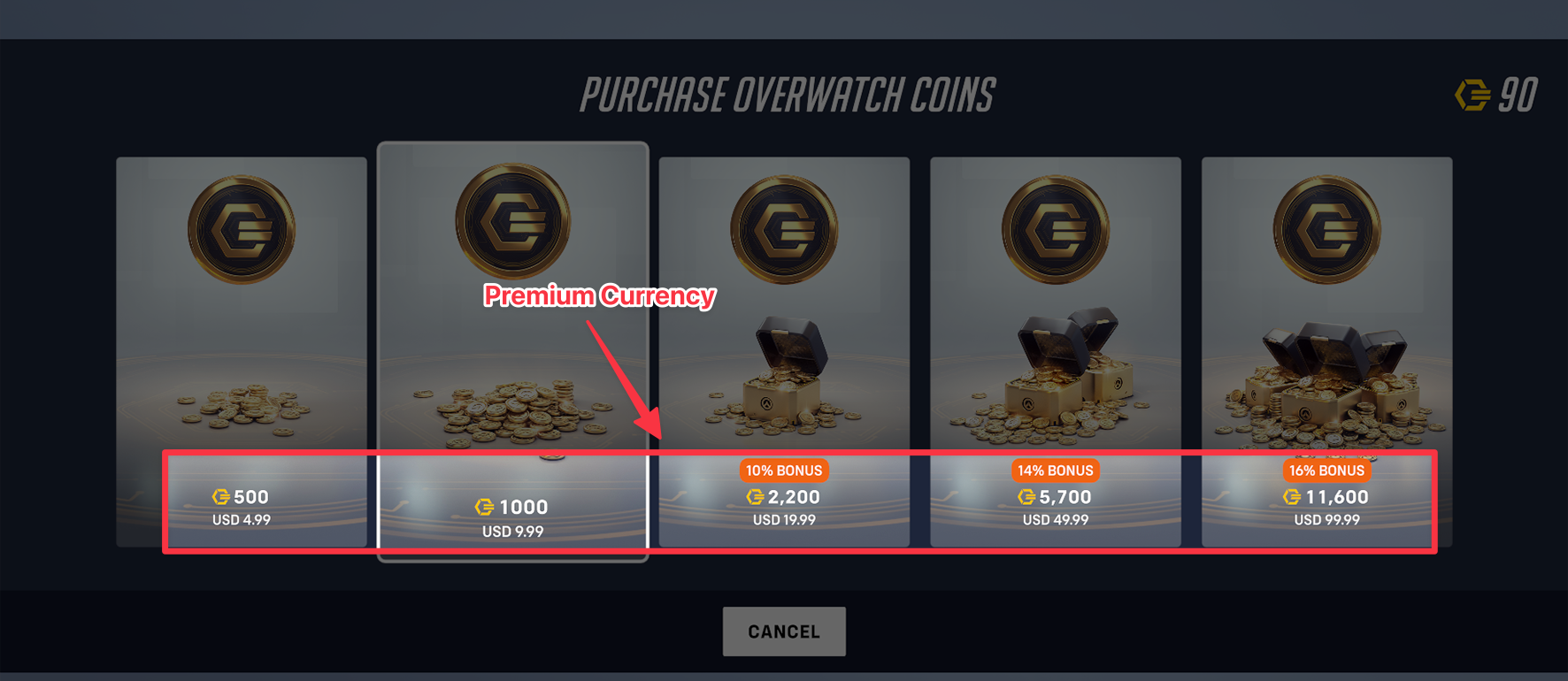}
        \caption{OW coins purchasable with real money (\textit{Premium Currency}). 
        }
        \Description{Purchasable OW coins (with real money). 
        }
        \label{fig:Coins-screenshot}
    \end{minipage}
\end{figure}

The addition of ``manipulative'' game mechanics during OW's \bmt has sparked active debates among the player community\footnote{See footnote 2.}. The first OW game (OW1) was released by Blizzard Entertainment in May 2016 as a B2P game and received critical acclaim for its smooth gameplay and vibrant art style~\cite{visual2016OW1}. Players can choose between 32 available characters and compete in several team-based combat scenarios. In OW1, players could earn \textit{lootboxes} as rewards by leveling up, winning their first daily game, winning in specific game modes, and completing holiday event tasks. Fresh content, seasonal events, and new characters were regularly added to keep players engaged. The second Overwatch game (OW2) was launched in October 2022 as a free-to-play (F2P) sequel to OW1, after OW1 servers shut down. OW2 retained the core game mechanics of OW1 with additional features, such as challenges (\autoref{fig:Challenges-screenshot}), battle passes (\autoref{fig:BattlePass-screenshot}), game shop (\autoref{fig:Shop-screenshot}), and new characters (\autoref{fig:character-screenshot})~\cite{Blizzard2022Welcome} (See \autoref{tab:glossary} in the Appendix for a glossary of in-game items). Without lootboxes, the primary rewards in OW2 come from the seasonal \textit{battle pass} and real money purchases from the \textit{game shop}. 


\begin{figure}[!t]
    \centering
    \begin{minipage}{0.48\textwidth}
        \centering
        \includegraphics[width=\textwidth]{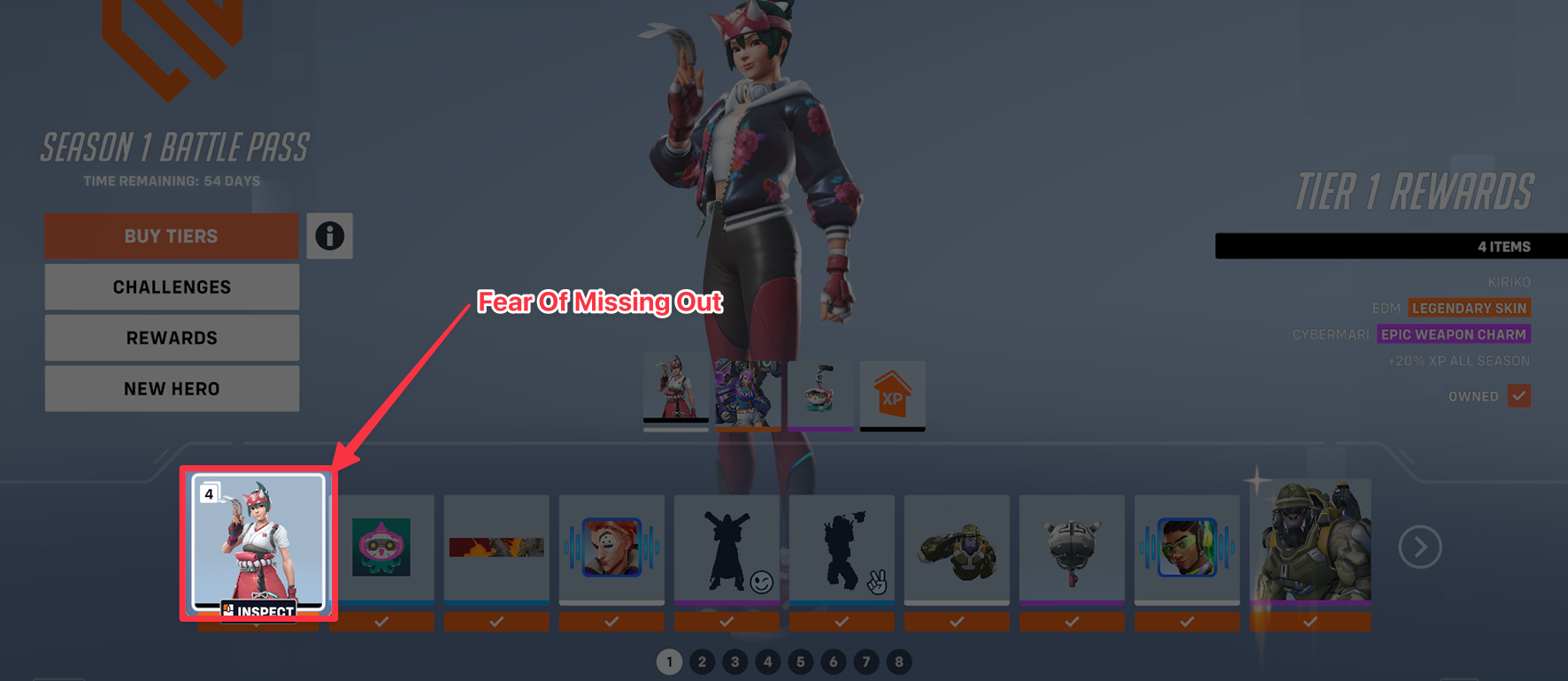}   
    \end{minipage}\hfill
    \begin{minipage}{0.48\textwidth}
        \centering
        \includegraphics[width=\textwidth]{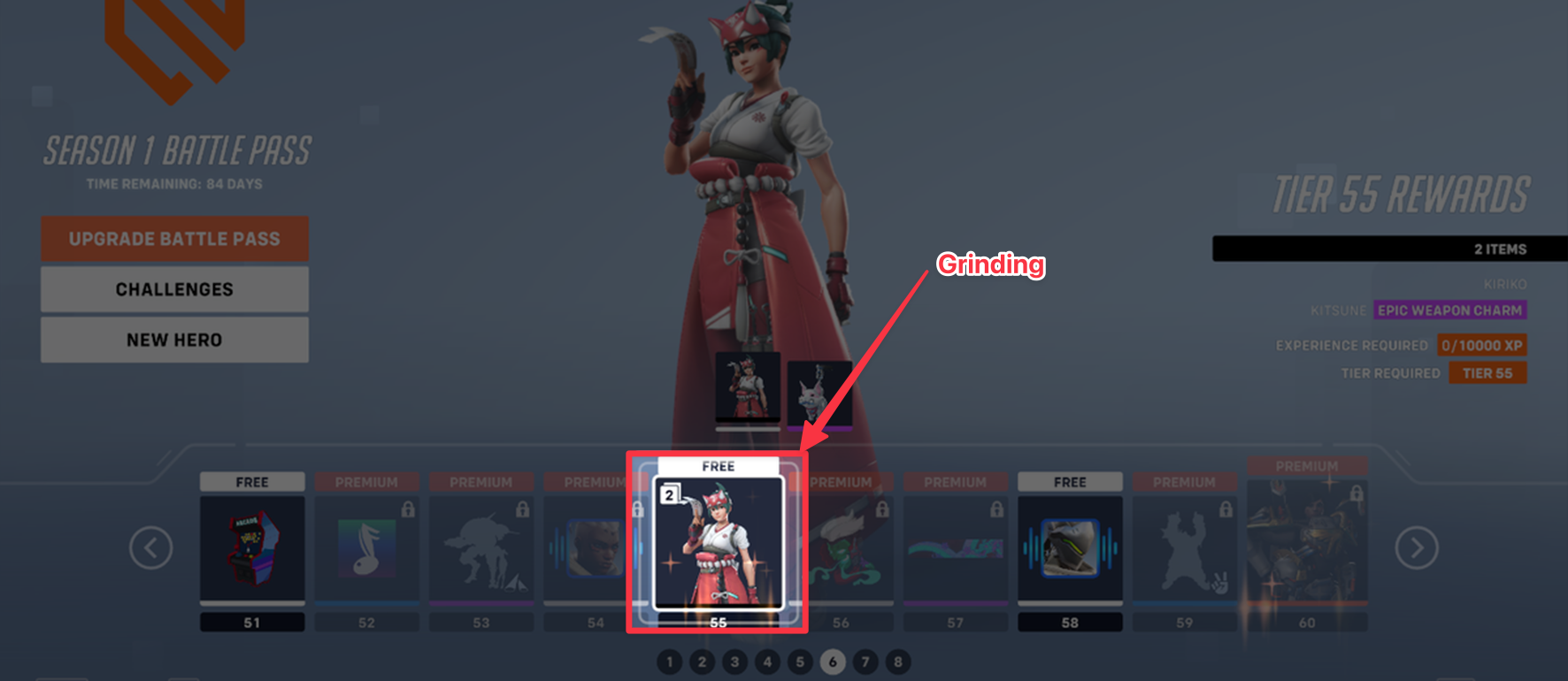}
    \end{minipage}
    \caption{The new unlockable character from OW2 season 1. Players who have purchased the premium \BP can unlock this character once they complete level 1. Other players can unlock this character once they complete level 55 (\textit{Grinding}). Players who miss this opportunity will need to complete character-specific challenges in later seasons, a much lengthier and more difficult path of obtainment (\textit{Fear of Missing Out}). 
    }
    \Description{The new unlockable character from OW2 season 1. Players who have purchased the premium \BP can unlock this character once they complete level 1. Other players can unlock this character once they complete level 55 (in season 1).
    }
    \label{fig:character-screenshot}
\end{figure}

\input{tables/Table-I}

Our research extends the Human-Computer Interaction (HCI) literature on deceptive design in games by examining deceptive design practices in a game's \bmt and their impact on players. However, we do not aim to attribute blame or criticism. Rather, we intend to learn from players who experienced the OW transition to understand the effects of deceptive design and to propose solutions for facilitating ethical game design going forward.

The results of our research make three main \textit{contributions}: First, we began with an OW2 game mechanics analysis and identified deceptive patterns in nine newly-introduced game design mechanics as a result of the game's \bmt. This analysis identified deceptive design practices implemented \textit{during} the transition to F2P and provides the foundation for how we studied player perspectives and experiences. Second, based on a thematic analysis of Reddit posts, we examined players' perceptions and experiences of deceptive game mechanics. The analysis provides evidence of how deceptive design practices can significantly affect the player experience, resulting in frustration, disappointment, and game abandonment. Third, from players' expectations and suggestions, we found alternative solutions for satisfying and non-manipulative game design practices, which would improve a game's design after an F2P transition.

%% file: tables/Table-I.tex
\begin{table}[!ht]
\centering
\caption{Summary of Game Mechanics of OW1, OW2, and other popular F2P games.}
\label{tab:game-mechanics-summary}
\resizebox{\textwidth}{!}{%
\begin{tabular}{@{}llllll@{}}
\toprule
\textbf{Game}$^a$ & \textbf{Publisher} & \textbf{Game Genre} & \textbf{Initial Release} & \textbf{Core Gameplay Mechanic} & \textbf{Auxiliary Mechanics} \\\midrule
Overwatch 1 & Blizzard  & First-person & 2016---2022 & Two opposing teams combat & Lootboxes*, currency, account-based level, hero \\
(OW1) & Entertainment & shooter (FPS) & & each other in achieving the & cosmetics, achievements, time-limited events, \\
& & & & goal (e.g., control an object, & badges, social features.\\ \cmidrule(lr){6-6}
Overwatch 2 & Blizzard & FPS & 2022---present & or escort a payload) & Battle pass-based levels and rewards*, currency*, \\
(OW2) & Entertainment & & & & new character*, character cosmetics*, game shop*, \\
&  & & & & daily/weekly/seasonal challenges, badges,\\
&  & & & & achievements, time-limited events, social features\\ \cmidrule(lr){2-6}
Fortnite & Epic Games & FPS & 2017---present & players scavenge for gear to &Currency*, battle pass-based levels and rewards*, \\
&  & & & defend themselves from other & account-based level, cosmetic items*, game shop*, \\
&  & & &  players or computer-controlled & daily/weekly/seasonal challenges, time-limited\\
&  & & & enemies. & events, social features$^b$\\  \cmidrule(lr){2-6}
League Of & Riot Games & Multiplayer& 2013---present & Two opposing teams combat to & Currency*, battle pass-based levels and rewards*, \\
Legends& & Online Battle& & destroy the other's base. & account-based level, new character*, cosmetics \\
& &  Arena (MOBA) & & & items*, game shop*, character enhancement\\
& & & & &  collectables, daily challenges, time-limited events,\\
& & & & &  social features$^c$\\ \cmidrule(lr){2-6}
Genshin & miHoYo & Role-Playing& 2020---present & Explorable open-world with & Currency*, gacha system (new characters and\\
Impact& &  Game (RPG) & & storylines, puzzles, dungeons, &  weapons)*, energy*, battle pass-based levels and \\
& & & &  home building, and fishing. & rewards*, account-based level, character cosmetics*,\\
& & & & & game shop*, daily/weekly/bi-monthly challenges, \\
& & & & & story/world/character quests, achievements, \\
& & & & & character and weapon enhancement collectables,\\
& & & & & time-limited events, social features$^d$ \\\bottomrule
\multicolumn{6}{l}{\begin{tabular}[c]{@{}l@{}}\textit{Note}. This table includes a list of representative game mechanics from popular F2P games.\\ * items purchasable with real money without additional efforts (e.g., open up lootboxes, complete the dungeon, grind for levels).\\
$a.$ We included the OW game series and three popular F2P games with more than 50,000,000 active players (as of January 5, 2024). \\ 
$b.$ game mechanics retrieved from Fortnite Wiki.~\url{https://fortnite.fandom.com/wiki/Fortnite_Wiki}\\
$c.$ game mechanics retrieved from League Of Legends Wiki.~\url{https://leagueoflegends.fandom.com/wiki/League_of_Legends_Wiki} \\
$d.$ game mechanics retrieved from Genshin Impact Wiki.~\url{https://genshin-impact.fandom.com/wiki/Genshin_Impact_Wiki}.
\end{tabular}} \\
\end{tabular}%
}
\end{table}

%% file: sections/02-RelatedWork.tex
We first present an overview of deceptive design and its integration in games. We then discuss how the F2P game business model motivates deceptive design practices. Finally, we discuss how OW game series provide an ideal research context to address gaps in existing literature. 

\subsection{Deceptive Design and Its Implementations in Games}


Introduced in 2010 by~\citet{Brignull2010deceptive}, deceptive design refers to design practices that distort or impair users' ability to make informed decisions, regardless of the designer's intentions~\cite{mathur2021makes,gray2023towards,EUDSA2022}. Those practices persuade users to engage in undesirable actions or make unfavorable decisions, causing negative consequences~\cite{mathur2019dark,Brignull2023book}. Research has explored its applications on websites~\cite{Brignull2023book,mathur2019dark}, mobile apps~\cite{lewis2014irresistible,fitton2019F2P}, and games~\cite{zagal2013dark,dillon2020digital}. For example, \citet{mathur2019dark} classified seven categories of deceptive patterns based on a large-scale analysis of shopping websites. Taking inspiration from~\citet{mathur2019dark}, \citet{Brignull2023book} identified eight exploitative design strategies and seven types of deceptive design patterns from practical examples. \citet{gray2018dark} later extended Brignull's classifications into a taxonomy with five major categories. Building up on~\citet{Brignull2023book} and \citet{mathur2021makes}, \citet{mildner2023defending} and \citet{mildner2023engaging} identified deceptive design in social networking services. In 3D environments, \citet{Greenberg2014proxemic} considered how deceptive design could be used in proxemic interactions to trigger unwanted interactions. \citet{hadan2024deceived} synthesized deceptive design in immersive extended reality technologies from the literature.

Within the context of games, deceptive design refers to \textit{patterns that cause negative player experiences with a positive outcome for game developer}~\cite{zagal2013dark}. Based on examples from over 20,000 mobile games, \citet{zagal2013dark} defined four main deceptive design categories. Taking inspiration from~\citet{zagal2013dark}, \citet{fitton2019F2P} classified six categories of monetization mechanisms in F2P mobile apps, and \citet{karlsen2019exploited}'s investigation with three mobile and web-based games found that these games predominately relied on grinding-and-reward system and play-by-appointment patterns. In addition, \citet{geronimo2020UI} investigated user perception of deceptive design in mobile apps and games following~\citet{gray2018dark}'s taxonomy. 

Many academic researchers and government agencies have also sought to define deceptive design using different terminology (e.g., \cite{zagal2013dark, lewis2014irresistible, California2020CPRA, DETOURACT2019}), or have attempted to find consensus among existing definitions (e.g., \cite{mathur2021makes,gray2023towards,roffarello2023defining}). For instance, \citet{mathur2021makes} consolidated deceptive design studies across fields and synthesized six attributes and various terms that have been used to define deceptive design, such as ``coercive,'' ``manipulative'', ``misleading,'' ``steering,'' ``trickery,'' and ``subvert user intent''~\cite{mathur2021makes, Brignull2010deceptive, gray2018dark,bosch2016tales}. \citet{roffarello2023defining}'s literature review identified eleven deceptive design patterns that contribute to users' attention capture. \citet{gray2023towards} developed a domain-agnostic ontology categorizing deceptive design into high-, meso-, and low-level patterns for easier access and adaptation of existing knowledge. 

While these classifications formed a foundation for analyzing deceptive design in OW2 in our study, we mainly adopted \citet{zagal2013dark}'s classification since it is the most cited classification that exclusively focused on games. Despite criticisms of its lack of ``empirical grounding,'' this classification articulates particular values (e.g., transparency) and player experiences (e.g., regret) that ``indeed form fruitful analytic or empirical starting points for tracing when and why particular game design decisions can become ethically questionable''~\cite[p.~2]{deterding2020against}. Our two-phased approach grounded our research in an empirical analysis of players' perceptions on Reddit, therefore addressing this limitation within the~\citet{zagal2013dark}'s classification. 

\subsection{Game Monetization that Motivates Deceptive Design}

In general, the video game industry employs three popular monetization models~\cite{massarczyk2019economic}: Pay-to-Play (P2P), Buy-to-Play (B2P), and Free-to-Play (F2P). P2P games require players to pay a monthly or yearly subscription fee, sometimes with an additional upfront purchase of the game, providing publishers and developers with a stable recurring revenue stream~\cite{massarczyk2019economic}. However, this model has lost popularity in recent years as players increasingly resist continuous fees~\cite{massarczyk2019economic}. B2P games, unlike P2P, allow players to buy a fully-developed game for a one-time cost~\cite{massarczyk2019economic}. Publishers often opt for \textit{micro-transactions}, offering additional Downloadable Contents (DLCs) and cosmetic items to generate more revenue~\cite{vanhatupa2011business, massarczyk2019economic}. However, players often argue that they should not have to spend more after buying the game for a complete experience~\cite{massarczyk2019economic}. 
In contrast to the B2P and P2P models, F2P games allow players to join for free but offer in-game items for purchase, serving as a revenue source to cover development costs~\cite{massarczyk2019economic}. ``Freemium'' is perceived as the best among the three, because it lowers player entrance barriers and increases developer and publisher profits~\cite{macinnes2002business, massarczyk2019economic, sanchez2022welfare}. Players can lose track of their spending through micro-transactions, ultimately spending more than anticipated~\cite{massarczyk2019economic}.

As the F2P games' revenue generation mainly depends on players' in-game spending, developers are incentivized to facilitate user motivation through game design, even using tactics to shape players' desires and behaviours~\cite{alha2014free}. Literature has studied the motivations behind player engagement, acquisition of in-game items, and in-game purchases (e.g.,~\cite{yee2006motivations,lehdonvirta2009virtual,toups2016collecting,hamari2017players}). However, such motivations may also be influenced by intentional game design choices~\cite{hamari2010game,paavilainen2013social}. In fact, numerous unethical, unfair, and deceptive tactics have been developed to leverage these motivations for the purpose of boosting game profits. For instance, \textit{predatory monetization} uses designs that obscure or delay the disclosure of the true long-term costs until players are committed either psychologically or financially~\cite{king2018predatory}. Examples include locking essential quality-of-life game aspects behind paywalls, and pay-to-win schemes~\cite{petrovskaya2021predatory}. Beyond monetization, games also use \textit{deceptive design} to maintain constant player engagement through temporal, social, and psychological tricks~\cite{zagal2013dark,roffarello2023defining,gray2018dark}. Various metrics have been used to measure the success, including daily/monthly active user (DAU/MAU), average revenue per user (ARPU), and players' lifetime value and estimated duration of interest in the game (LTV)~\cite{fields2011social}.

\subsection{The Role of Player Perception in Analyzing Deceptive Game Design}

Deceptive designs exploit people's internal biases, such as \textit{bounded rationality}, \textit{loss aversion}, \textit{temptation}, and \textit{mindlessness}; and external influences such as \textit{social influence} that contribute to the complexity of human decision-making~\cite{simon1997models,waldman2020cognitive,thaler2009nudge}. The effectiveness of deceptive design varies depending on users' literacy in recognizing such patterns~\cite{zagal2013dark,geronimo2020UI,luguri2021shining}. 
For instance, \citet{geronimo2020UI} uncovered users' unawareness of deceptive design's existence in mobile apps. \citet{luguri2021shining} demonstrated that subtle deceptive designs are more easily overlooked, especially by less-educated users. Players' perceptions also matter, as some may feel deceived and angry but others might appreciate the guidance~\cite{zagal2013dark,frommel2022daily}. For example, \textit{Daily Tasks} could be seen positively by some players (for accomplishment feelings), while others may view them negatively due to `Fear of Missing Out' (FOMO) and the obligation to play~\cite{frommel2022daily}. \textit{Grinding} might be found worthwhile when associated with variations, achievements and progress~\cite{karlsen2019exploited,zagal2013dark}.  

In addition to classifying deceptive design, research has also investigated its potential impact from users' perspectives. For instance, \citet{maier2019dark} investigated user perceptions, experiences, and awareness of deceptive design, revealing a resigned attitude among users. Users blamed businesses for the occurrence of deceptive design but simultaneously realized their dependency on its services, making users hard to avoid deceptive designs~\cite{maier2019dark}. Building on this, \citet{m2020towards} identified five factors influencing users' susceptibility to deceptive design: the frequency of occurrence, perceived trustworthiness, level of frustration, misleading behavior, and physical UI appearance. Furthermore, \citet{fitton2019F2P}'s deceptive design framework included designs that ``interrupted'' or ``annoyed'' users during their interaction with the mobile apps and games. From a player's perspective, \citet{karlsen2019exploited} uncovered the mixed effects of temporal patterns. \citet{gray2021enduser} argued that users' experiences and perceptions of manipulation can facilitate identifying problematic practices that may not be strictly illegal. The systematic review by~\citet{gray2023mapping} recommended incorporating explicit human-in-the-loop techniques in studies that focusing on deceptive design detection to overcome the lack of detectability of many pattern types. Our research methodology echoes these recommendations and conclusions from prior studies by incorporating an analysis of player experiences and perceptions on the deceptive design in OW2. 

\subsection{Research Gaps and Research Approach}

Although many P2P and B2P games are transitioning to F2P (e.g.,\textit{Team Fortress 2}~\cite{rizani2020analysis}, \textit{Counter-Strike: Global Offensive}(CS:GO)~\cite{rizani2020analysis}, \textit{Overwatch}~\cite{newham2022consequences} ), few studies have explored its effects on players. Existing literature has primarily focused on game publishers. For example, \citet{newham2022consequences} analyzed \textit{Overwatch}'s design against existing design frameworks and concluded that the transition from B2P to F2P can be successful and beneficial to the publisher. \citet{rizani2020analysis} studied \textit{CS:GO} and \textit{Team Fortress 2} and identified increased player numbers and activity after the \bmt. 
However, the expectations of players 
and potential strategies to avoid problematic design practices and ensure a
satisfactory game \bmt for both publishers and players remain unexplored. The perspective of players is a valuable resource for understanding the effects of game tactics for game designers~\cite{petrovskaya2021predatory, Shokrizade2013f2p, lin2011cash}. 
\minor{For this reason, we ask our \textit{RQ1: How do players perceive the role of deceptive game mechanics resulting from the business model transition?}} 

Previous research has identified deceptive designs in games (e.g., \citet{zagal2013dark, geronimo2020UI, fitton2019F2P}). However, the use of deceptive design practices and their effect on players during \bmt have been largely overlooked. We fill this gap by investigating how deceptive design practices in a B2P-to-F2P game \bmt affect players' in-game experience and perceptions. \minor{We ask \textit{RQ2: What elements contribute to a satisfying and non-manipulative player experience?}} We focus on studying the \textit{Overwatch} series because this game recently undergone a \bmt, which evoked extensive discussion among its players\footnote{See footnote 2.}. We began by analyzing game mechanics that newly introduced into OW2 as a result of \bmt and identifying those that exhibit deceptive design characteristics~\cite{zagal2013dark}. Then, we conducted a thematic analysis of players' reviews on Reddit to capture game player interactions, perceptions, and experiences that contribute to their feeling of being manipulated. We combined our game mechanics analysis results with data from Reddit. Our two-step process addressed the subjective nature of (and lack of empirical grounding for) calling an interface ``deceptive''~\cite{gray2018dark,luguri2021shining,zagal2013dark}. It confirmed the accuracy and validity of our results and gave us a better understanding of how this ``deceptiveness'' spreads in the game. Player perceptions are valuable indicators of optimal game usability and player experience~\cite{nacke2018games}. Hence, this approach also allowed us to go beyond game UI-level deceptive design and identify where players felt being manipulated during their interaction sequences, regardless of the designers' intentions.

%% file: sections/03-Methodology.tex
We study the role of deceptive design in a game’s business model transition from players’ perspectives to identify ways to reduce deceptive design practices and improve user experience during this transition. Following methodology from literature~\cite{mildner2023engaging,karlsen2019exploited,geronimo2020UI,king20233d}, we took a two-phase method: 1) an analysis on game mechanics newly introduced into OW2 from the \bmt and identify those exhibit deceptive design characteristics, and 2) a thematic analysis on Reddit reviews to understand how these potentially deceptive designs were experienced and perceived by the players. The first method analyzes the game as ``structure'' and the second analyzes the game as play experiences~\cite{karlsen2019exploited}. 

\subsection{Game Mechanics Analysis}
\label{subsec:UI-analysis-method}

We conducted an analysis of newly introduced game mechanics in OW2 during the \bmt that did not exist in OW1. We did not focus on the potential deceptive game mechanics already existing in OW1 as players long-time exposure may have led to habituation of deceptive design and reduced negative responses. Then, we extracted the game mechanics that exhibit deceptive design characteristics~\cite{zagal2013dark}. Game mechanics are the rules and systems that govern players' interactions, while game design patterns are recurring design structures or elements that implements game mechanics in games~\cite{adams2012game}. 

\subsubsection{Game Mechanics Data Collection}

We recorded a 30-minute video, where the first author interacted with all OW2 in-game interfaces. From the video, multiple screenshots of each game mechanic were taken to demonstrate players' interaction sequence. Since OW1 was discontinued in October 2022~\cite{Blizzard2022Welcome}, we obtained screenshots of OW1 game mechanics from the \textit{Game UI Database}\footnote{Game UI Database. \url{https://www.gameuidatabase.com/gameData.php?id=1341}}, \textit{Interface In Game}\footnote{Interface In Game. \url{https://interfaceingame.com/games/overwatch/}}, and gameplay streams on \textit{Youtube}\footnote{\url{https://www.youtube.com/}} and \textit{Twitch}\footnote{\url{https://www.twitch.tv/directory/gaming}}. In addition, a researcher with 5-year experience playing OW1 \& OW2 reviewed the screenshots to ensure that our screenshots inclusively captured all commonly used OW1 and OW2 game mechanics. Since our study focuses on studying differences between OW1 and OW2, we \textit{excluded} all the game mechanics that the two games shared in common. The resulted 62 screenshots were then uploaded to Dovetail\footnote{Dovetail --- Qualitative Coding Platform.~\url{https://dovetail.com/}} for the next step. 

\subsubsection{Identifying Candidate Deceptive Game Mechanics}

While various studies have explored deceptive design in games~\cite{fitton2019F2P, karlsen2019exploited, geronimo2020UI}, a precise and standardized taxonomy for this type of game design does not exist. Our game mechanics analysis adopted a deductive approach, using the existing deceptive design classification from~\citet{zagal2013dark} (patterns in each category are explained in~\autoref{tab:game-deceptive-design} in the Appendix):

\begin{itemize}
    \item \textbf{Temporal Patterns} ``cheat'' players out of their time, making them spend more/less time than expected.
    \item \textbf{Monetary Patterns} deceive players into spending more money than anticipated.
    \item \textbf{Psychological Patterns} dupe players with psychological tricks.
    \item \textbf{Social Patterns} use players' relationships with friends and family to benefit the game.
\end{itemize}

Five researchers evaluated the 62 screenshots in a collaborative review session. During this review session, we labelled OW2 screenshots based on deceptive design patterns from four categories: \textit{temporal patterns}, \textit{monetary patterns}, \textit{social patterns}, and \textit{psychological patterns}~\cite{zagal2013dark}. Following the approaches of prior research~\cite{geronimo2020UI,mildner2023engaging}, we coded screenshots based on perceived problems in the game mechanics---those that work against players' ``best interests'' and could benefit the game and the publisher---rather than focusing on designer intent. This approach allowed us to identify deceptive design that emerged from both deliberate manipulation by designers and unintended yet potentially harmful design outcomes. The five researchers iteratively reviewed and discussed the labels until they reached full agreement on their validity and fit. \UImethod{In this process, we started with the core interfaces of the game mechanics and expanded on our discussion to other related interfaces that players could encounter during their interactions. We incorporated a specific example of our labelling process in~\autoref{sec:app-sample-coding}}. In total, we identified nine potentially deceptive OW2 game mechanics that demonstrated the characteristics of 12 deceptive patterns. 
\autoref{tab:UI-analysis} shows a summary of these OW2 game mechanics. These game mechanics served as the starting point for our investigation on players' perceptions and experiences.

\subsection{Thematic Analysis of Player Comments on Reddit}
\label{subsec:reddit-analysis-method}

For the second part of our research, we analyzed players' perceptions and experiences of the nine potentially deceptive game mechanics we identified previously from two subreddits: \textit{r/Overwatch} and \textit{r/Overwatch2}. This step allowed us to mitigate the subjectivity in our game mechanics analysis, and determine
instances where players felt manipulated during their interaction sequences. 
We intentionally excluded Blizzard forums moderated by the game publisher due to the likelihood of censorship or removal of controversial topics about the game or the company.\footnote{This is evident by several Reddit posts where creators  assert that they were "silenced" for discussing their experience with Blizzard support.} We selected the two subreddits because of their large number of active followers. As of April 2023, \textit{r/Overwatch} had more than 4 million followers and \textit{r/Overwatch2} had more than 115,000 followers. Both were ranked as the top 1\% biggest subreddits among the Reddit community.

\begin{figure}[t]
  \includegraphics[width=\textwidth]{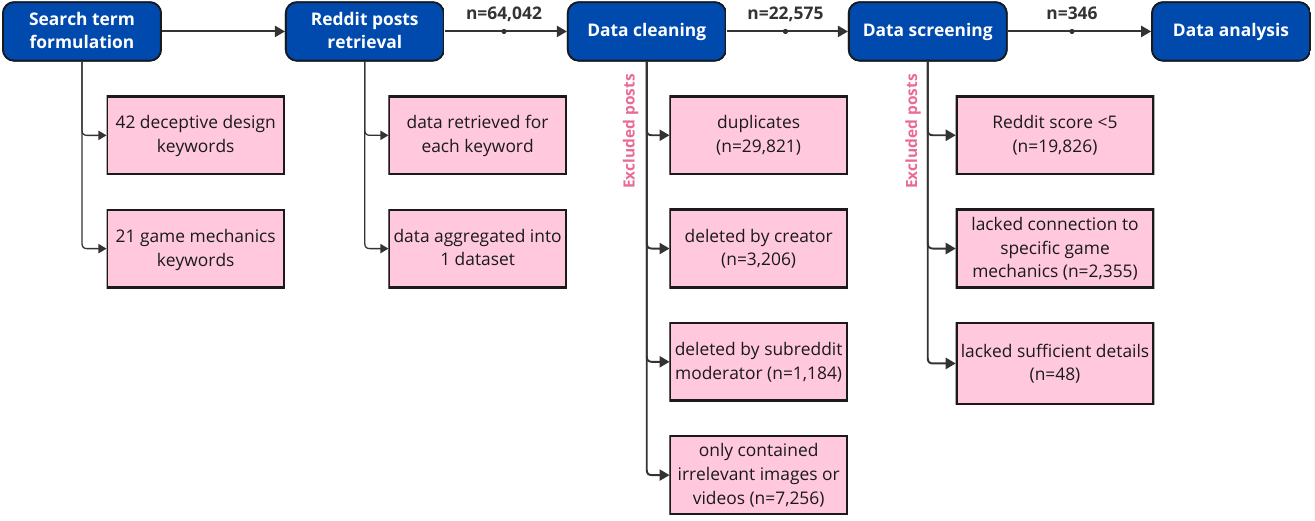}
  \caption{The flowchart shows our Reddit data collection process, from search keywords to data cleaning, screening, and final thematic analysis sample.}
  \Description{The flowchart shows our Reddit data collection process, from search keywords to data cleaning, screening, and final thematic analysis sample.}
  \label{fig:Reddit-flowchart}
\end{figure}

\subsubsection{Subreddit Data Collection}
\autoref{fig:Reddit-flowchart} summarizes our data collection process. We begin by identifying keywords related to deceptive designs (as discussed in~\autoref{sec:related-work}) and keywords describing the nine potentially deceptive game mechanics from our analysis in~\autoref{subsec:UI-analysis-method}. Then, we created different variations of the keywords. For example, variations of the keyword ``manipulate'' included ``manipulating'' and ``manipulative.'' To increase the relevancy of the results, we performed test searches iteratively on \textit{r/Overwatch} and \textit{r/Overwatch2}. We arrived at a final set of \textit{63 search keywords} (listed in~\autoref{sec:app-reddit-keywords}) that were used for data retrieval using a Python script with the Reddit API (PRAW\footnote{PRAW 7.7.1 Documentation.~\url{https://praw.readthedocs.io/en/stable/}}).  

Data for each keyword were retrieved from Oct 4, 2022 to April 27, 2023 and were aggregated into a large dataset, resulting in a total of 64,042 candidate posts. After eliminating duplicates, blank posts deleted by their creators or the subreddit moderator (i.e., only contain ``[deleted]'' or ``[removed]'' as content), and those that only contained irrelevant images or videos, we retained 22,575 posts. To avoid data over-saturation and to maximize our efforts for posts with valuable content, we further excluded posts with a $<5$ Reddit score\footnote{Reddit score, or ``submission score,'' is the difference between upvotes and downvotes on a Reddit post. For instance, if five users upvote and three users downvote a post, its score would be 2. See: \url{https://www.reddit.com/wiki/faq/}}. We also excluded posts that expressed creators' general feeling of being manipulated or deceived \Postmethod{from their interaction with other players (e.g., manipulative teammates, deceptive gameplay strategies of opponents, or unsatisfactory games manipulated by intentional player grouping)} without connection to specific game mechanics, and posts that lacked sufficient details for analysis. Our final dataset contained $n=346$ highly relevant and highly rated posts that appeared in the searches using both deceptive design-relevant and game mechanics-relevant keywords.

\subsubsection{Qualitative Data Analysis}
We analyzed the Reddit posts on Dovetail. 
Three researchers analyzed the Reddit posts in an iterative process using inductive \textit{Thematic Analysis}~\cite{clarke2015thematic}.During this coding process, we did not rely on the keywords used for data retrieval, but rather focused on the emergent themes within the content itself.
First, the researchers browsed the posts to familiarize themselves with the data. Next, three researchers independently open-coded 10\% of the posts (35 posts). In a subsequent meeting, they discussed the open codes, resolved conflicts, and formulated an initial codebook. The meetings with three coders continued weekly for 10 weeks to refine the codebook until all posts were coded, no new codes emerged, and all conflicts were resolved. Finally, the researchers collaboratively developed themes and sub-themes, shown in~\autoref{tab:RQ3} in Appendix. 

\subsection{Ethical Considerations}
\label{subsec:ethical-considerations}

Our game mechanics analysis was performed by researchers who are not considered ``human subjects''~\cite[p.~14]{TCPS2018}. We restricted our data collection to two subreddits that were publicly accessible, registration-free, and are considered ``publicly available'' material~\cite[p.~17]{TCPS2018}. To help protect the anonymity of the post creators, we did not collect personally identifiable information. In presenting our results, we removed the specific subreddit name and paraphrased all direct quotes in such a way that the original post and the creator are not easily traceable.

%% file: sections/04-Results.tex
We organized our results as follows. First, we present the findings of our analysis on OW2 game mechanics and ground these findings in our thematic analysis of players' review on Reddit. Second, we go beyond game design mechanics and provide insights from players' perceptions of how the ``feeling of being manipulated'' resulted from the \bmt and the game publishers' business practices. In~\autoref{sec:discussion}, we discuss our findings within broader-scope deceptive design taxonomies (e.g.,~\cite{mathur2021makes,gray2023towards}) and draw implications for game designers, publishers, and researchers.

\subsection{Deceptive Game Mechanics and Players' Experiences and Perceptions}
\label{subsubsec:RQ1}
 
To answer our RQ1, we identified nine OW2 game mechanics introduced during the \bmt that exhibited characteristics of 12 deceptive design patterns~\cite{zagal2013dark}. A game mechanic might demonstrate characteristics of multiple deceptive patterns, and a deceptive pattern might be integrated in multiple game mechanics. A short description of these game mechanics and the corresponding deceptive patterns can be found in~\autoref{tab:UI-analysis}. For clarity in the paragraphs below, game mechanics are set in monospace font (e.g., \varname{\BP}), deceptive patterns are shown in title-case \textit{italics} (e.g., \textit{Grinding}), descriptions directly extracted from the game are placed within quotation marks, and excerpts from Reddit reviews are set in \textit{italics} text and quotation marks. 

\input{tables/Table-II}

\subsubsection{Time-sinking experiences formed by temporal patterns.}
Throughout our analysis, we found multiple game mechanics implemented temporal deceptive patterns to maintain player engagement, encouraging them to spend more time than they would have otherwise~\cite{zagal2013dark}. For example, \varname{Daily/Weekly/Seasonal Challenges} and \varname{\BP} demonstrated the characteristics of \textit{Grinding} and \textit{Infinite Treadmill}. As seen in example screenshots in~\autoref{fig:BattlePass-screenshot} and~\autoref{fig:Challenges-screenshot}, these game mechanics require continuous play daily, weekly, and seasonally to complete the challenges (e.g., ``earn 10 eliminations without dying''). Players' progress is reset periodically, forcing them to continue playing to accomplish the never-ending goals. Within the \varname{\BP}, a \varname{New Character} is available every two seasons (see~\autoref{fig:character-screenshot}). Players without a premium \varname{\BP} must grind to a particular tier (e.g., tier 55 in season 1) to obtain the character before the season ends, or they must complete seven special challenges afterward. Thus, this game mechanic exhibited the characteristics of \textit{Grinding}. In addition, \varname{Double XP Weekend} and the \varname{PACHIMARCHI Event}~\cite{Pachimarchi2023OW} are \textit{Playing-by-Appointment} patterns that occur only during a specific time, compelling interested players to play during those time frames.

\begin{figure}[t!]
  \includegraphics[width=0.7\textwidth]{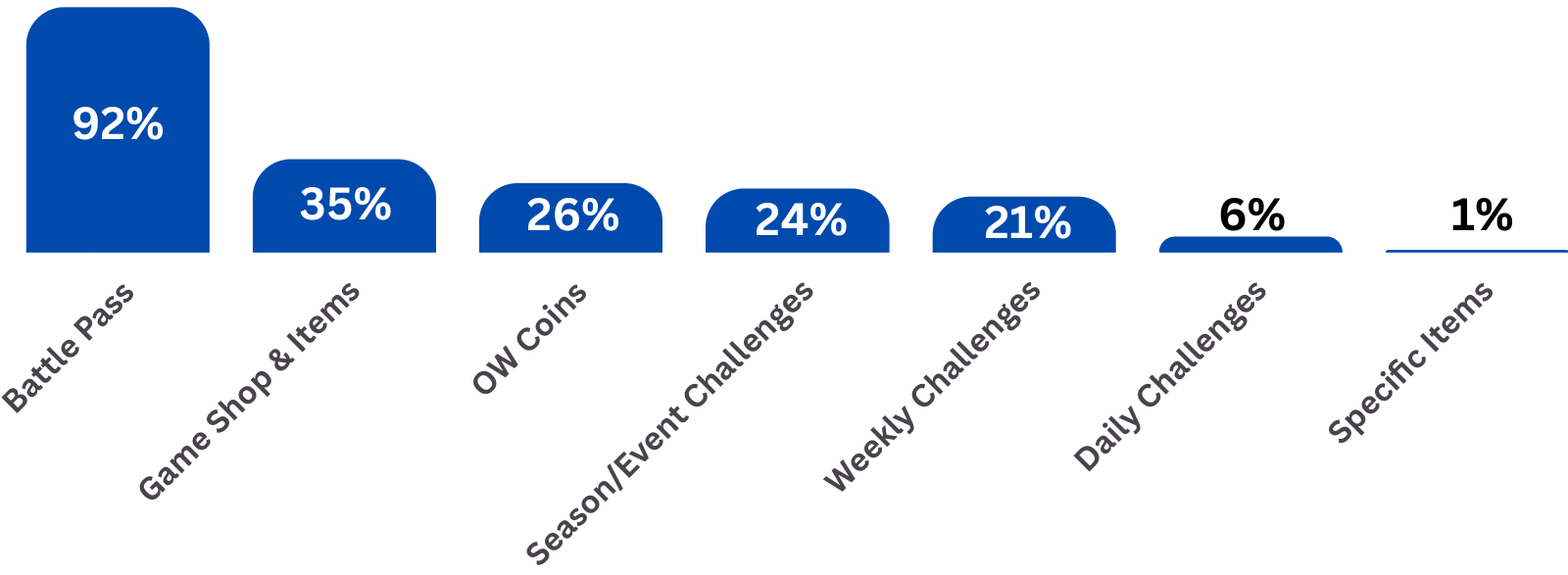}
  \caption{Percent of Reddit posts that mentioned each OW2 game mechanic.}
  \Description{Percent of Reddit posts that mentioned each OW2 game mechanic.}
  \label{fig:OW2-game-mechanics-percent}
\end{figure}

Our analysis of Reddit posts confirmed the impacts of these game mechanics. As presented in~\autoref{fig:OW2-game-mechanics-percent}, the majority (92\%) of the posts discussed issues regarding the \varname{\BP} system and the \varname{New Character} it offers. In these posts, players called the current \varname{\BP} system the worst among other F2P titles like Fortnite, Destiny 2, and Call of Duty. Several posts explicitly discussed about the \textit{``extensive 56-hour grinding''} required to finish the 80 tiers in Battle Passes. Players without a premium \varname{\BP} perceived having to grind to \varname{\BP} tier 55 (in Season 1) to get a new character, a \textit{``core game element,''} as not acceptable. Furthermore, 24\% posts mentioned problems players have had with \varname{Seasonal Challenges}, with a focus on the Halloween event, the first event after the \bmt. Twenty-one percent of posts mentioned \varname{Weekly Challenges'} coin-attainability and difficulty, and 6\% noted \varname{Daily Challenge} issues. Players felt being forced to spend a bit time a day on the \varname{Daily Challenge} to earn experience points (XPs) for the Battle Pass. Although \varname{Weekly Challenges} allowed the accumulation of OW coins, casual players perceived that an estimated 32 weeks of constant grinding to get a Legendary skin was a \textit{``huge ask''} for their time. \varname{Seasonal Challenges} further added to players' frustration by requiring them to play unfamiliar roles or win games in special game modes. Overall, players felt trapped in these time sinks and dreaded the never-ending and tedious tasks, which made the game feel like a chore rather than a hobby to them. We did not find posts discussing the \varname{PACHIMARCHI Event} or \varname{Double XP Weekend} possibly because their rare occurrence may have resulted in less player discussions.

\begin{quoting}

    ``Why limit currency rewards to weekly challenges? People don't have the time to complete most of them, and even finishing just one is too much for casual players.'' (Reddit post 192)

\end{quoting}

\subsubsection{Money-sinking experiences formed by monetary patterns.}
Our analysis identified multiple monetary deceptive design patterns within OW2 game mechanics that aim to get players spending more than they intend~\cite{zagal2013dark}. For example, \varname{\BP} incorporates characteristics of \textit{Anchoring Tricks}, \textit{Recurring Fee}, and \textit{Pay-to-Skip}. The premium \varname{\BP} fee potentially encourages players to grind it to get their money's worth before the season ends (approximately nine weeks~\cite{OW2seasonduration}). OW2 also allows purchasing \varname{\BP} tiers for players who do not want to spend time grinding and those who cannot finish it in time. In addition, within 80 \varname{\BP} tiers, players are rewarded with 80 items for the \$10 USD premium fee, which is seemingly cheaper than buying 80 items from the game shop. Furthermore, the \varname{In-Game Shop} (see~\autoref{fig:Shop-screenshot}) uses \textit{Anchoring Tricks}, \textit{Artificial Scarcity}, and \textit{Waste Aversion}. The shop inventory refreshes periodically to create fake scarcity, encouraging impulse purchases. Bundles are priced differently than purchasable OW coins (see~\autoref{fig:Coins-screenshot}), resulting in leftover coins and leading players to buy more coins to avoid wasting unused ones. In OW2, we consider \varname{OW Coins} as a deceptive pattern because coins can now be purchased with real money. This \textit{Premium Currency} establishes an exchange rate with real money, hiding the true prices of items.

From our Reddit player review analysis, we found evidence that premium \varname{\BP} buyers felt bound by it to grind as much as possible to avoid wasting the premium fee. Although \varname{\BP} includes rewards for free players (see~\autoref{fig:BattlePass-screenshot}), many posts indicated that players felt forced to purchase the premium to feel like they have made progress in-game. \varname{Halloween Event} was also perceived as less exciting since event-themed items can only be purchased from the \varname{Game Shop} instead of being rewarded through gameplay and lootboxes, like in OW1. 

In addition, 35\% of Reddit posts discussed the issues surrounding \varname{Game Shop} and \varname{In-Game Items}, including pricing and promotion. Twenty-six percent of posts further discussed players' frustrating experiences with \varname{OW Coins}. Many called the \varname{Game Shop} prices \textit{``insane,''} compared to other F2P games. Most items cost tremendously more than in OW1. From example, OW1's 250-coin Highlight Animations costs 750 coins in OW2, a 200\% cost increase. Players perceived it as disrespectful when they had to pay \$19 USD for a skin that was once free in OW1's lootboxes. One post claimed that the OW2 store page intentionally swapped the ``ok'' and ``cancel'' buttons from OW1, resulting in an accidental purchase. This player experience shows that \varname{Game Shop} incorporates the characteristics of \textit{Accidental Purchases}, which was not identifiable from a mere game mechanics analysis. Blizzard's 2022 fourth-quarter financial report proves these tactics financially successful, disclosing that the sales of OW digital products were at the highest tier to date.\footnote{Blizzard 2022 Financial Report.''~\url{https://investor.activision.com/news-releases/news-release-details/activision-blizzard-announces-fourth-quarter-and-2022-financial}} 

\begin{quoting}

    ``Shop prices are ridiculous. I don't understand why they don't make skins more affordable and accessible instead of charging players \$20 for each.'' (Reddit post 322)

\end{quoting}

\subsubsection{Social consequences formed by social patterns.} 

Our game mechanics analysis found that \varname{Daily/Weekly/Seasonal Challenges}, the \varname{\BP} and the \varname{New Character} further demonstrated the characteristics of \textit{Fear of Missing Out} (FOMO). However, insights from our analysis on Reddit posts revealed that players were discouraged from completing these challenges and the \varname{\BP} because of their perceived monotony and difficulty, with underwhelming rewards in return. For instance, several posts mentioned that the restriction of earning XP for only the first three \varname{Daily Challenges} impeded players' grind progress and increased frustration. Currently, \varname{Weekly Challenges} only yield 60 \varname{OW Coins}, which players considered \textit{``pathetic''} and \textit{``abysmal''} compare to the price of \varname{Game Shop} items. The \varname{Halloween Event Challenge} rewards were limited to weapon charms, voice lines, and sprays (see~\autoref{tab:glossary} in Appendix). However, in OW1, players could earn skins, which were frequently considered the \textit{``only thing that made OW1 so amazing.''}

\begin{quoting}

    ``Tedious challenges offer little reward for the effort required.'' (Reddit post 278)

\end{quoting}

Moreover, some posts discussed the peer pressure experienced by players in obtaining the \varname{New Character}. They felt pressure when their team required them to play it against the opposing team. Combining approaches like thematic analysis reveals additional layers of these manipulative practices, especially those rooted in \textit{Social Obligation/Guilds}. 

\begin{quoting}

    ``I can easily foresee a time when my team would blame me of not investing enough money or time to unlock a character.'' (Reddit post 279)

\end{quoting}

\subsubsection{Psychological impacts formed by psychological patterns.} 

Several OW2 game mechanics used psychological tricks~\cite{zagal2013dark}. For example, we identified the characteristics of \textit{Aesthetic Manipulations} from \varname{\BP}. Upon reaching tier 80 (see~\autoref{fig:BattlePass-screenshot}), premium players are rewarded with a mythic skin with customizable attire and unique audio effects. This exclusive mythic skin stands out from other skins in OW2 and cannot be obtained elsewhere, making the premium more tempting for players. In addition, the \varname{Daily/Weekly/Seasonal Challenges} and the \varname{\BP} also demonstrated the characteristics of \textit{Endowed Progress} by creating an artificial goal that players cannot easily quit.  

On the contrary, our analysis of Reddit posts revealed opposite findings. Players on Reddit were primarily concerned that the game lacked good rewards in compensating their time, money and cognitive investments in gameplay. They deemed the rewards for each \varname{\BP} tier \textit{``random''} and \textit{``unsatisfactory,''} compared to the grind. Some items (e.g., skins, emotes, victory poses, voice lines, and highlight intros) (see~\autoref{tab:glossary} in Appendix) were exclusive to specific characters. Given that OW2 had more than 30 characters available, receiving a \varname{\BP} item for a rarely played character felt pointless and unrewarding. The \varname{\BP} (in Seasons 1\&2) did not provide \varname{OW Coins} for players to buy items outside the pass, forcing them to submissively accept rewards identical to all other players, limiting their ability for personalization and self-expression.

\begin{quoting}

    ``It's frustrating that there's nothing worth grinding for underwhelming rewards.'' (Reddit post 388)

\end{quoting}

Moreover, players who obsessed with completing item collections felt unfulfilled in game. While the \textit{Complete the Collection} deceptive design~\cite{zagal2013dark} was also used in OW1, our analysis of Reddit posts revealed that the current OW2 reward system exacerbates its psychological impact, leading players to grind up to 326 years to earn enough coins or to spend up to \$10,200 USD to complete collections. 

\begin{quoting}

    ``Completing all hero item collections require players to grind for 17k+ weeks to get the coins necessary. Do they [the publisher] expect this game to last until [year] 2350?'' (Reddit post 90)

\end{quoting}

\subsection{How do players perceive the business model transition?}
\label{subsubsec:RQ2}

\subsubsection{Predatory marketing strategies}

Beyond specific game mechanics, many Reddit posts reported the game publisher's use of predatory marketing tactics---starting with high prices and dropping them to appear to respond to community feedback and to increase the players' impulse to purchase. As players feel hopeless to obtain an item through gameplay, they will spend more money, considering the price drops as a \textit{``gain.''} While many posts called the present system \textit{``greed[y]''} and claimed that players should be able to choose their own prices as consumers, some raised concerns that players are brainwashed to believe that it is reasonable for a game to implement deceptive tactics to generate revenue. For example, the game publisher emphasized the \textit{``tremendous value''} of the \varname{\BP} rewards, which they claimed could have been sold directly for more revenue. 

\begin{quoting}

    ``Don’t be manipulated! As consumers, we should determine the costs we're willing to bear.'' (Reddit post 20)

\end{quoting}

Moreover, the \varname{Game Shop} was also criticized for offering \textit{``fake''} bundles with items that were never available individually, forcing players to purchase the full bundle for one desired item. \textit{``Fake''} discounts were also offered on items never sold at full price. According to players on Reddit, \textit{``false claims and misleading impressions''} breached Consumer Protection Acts in many countries. 

In addition, the game publisher's lack of clarity about the value of \varname{OW1 Coins} in OW2 prior to the \bmt also confused and frustrated players. Many posts expressed players' fear of losing their OW1 coins in OW2 and they were pressured spend all their coins before the \bmt. These players criticized the game publisher for purposefully leaving ambiguity to encourage players wasting their OW1 savings so they had to spend real money on OW2 items. Some posts reported that OW1 coins only had limited utility in OW2 and could only be used to buy default items---not even items previously purchasable in OW1---making OW1 coins practically became worthless. Players saw this transition as \textit{``theft''} because they earned the coins through hard gameplay or purchased OW1 lootboxes. Overall, while players acknowledged that the long-term sustainability of the game depends on a good revenue structure, the current OW2 business strategies were perceived as problematic and harmed player experience. 

\begin{quoting}

    ``They seem deliberately vague. (OW1) Coins are a scam transfer. The shop was replaced with one which 
    the (OW1) currency they provided is not practically redeemable.'' (Reddit post 187)

\end{quoting}

\subsubsection{Disappointment outweighed enjoyment}

Players found that the changes in the core game mechanics enhanced OW2's competitiveness and overall enjoyment. As noted in many posts, transitioning to a F2P model expanded the player base, shortened game queue times, and allowed gameplay with friends and family. While players acknowledged the current game's positive impact on fostering social connections, auxiliary mechanics such as rewards and progression systems, which players found as significant to their motivation, resulted in negative perceptions and experiences. Many posts mentioned losing interest in cosmetics and rather sticking with default because of unwillingness to buy unfairly priced and lacklustre items with the hope to save OW2 coins for better items in the future. Some players even feel embarrassed using skins during gameplay because the \$19 USD purchase price diminished the perceived \textit{``coolness.''} 

\begin{quoting}

    ``Lets be honest, nobody envies the loser who spent money on a bad deal.'' (Reddit post 9)

\end{quoting}

The lack of good-quality rewards in the game deterred players and lowered their enthusiasm for gameplay. Many posts argued that the game should not be dubbed F2P with its existing reward system and F2P should not mean free players only get the crumbs from the plate.

\begin{quoting}

    ``With all the locked heroes and cosmetics, I can't see myself joining the game that pushes me to spend money. Nobody wants to play a game that seems to hate its players.'' (Reddit post 322)

\end{quoting}

Our analysis included Reddit posts from old OW1 players and those who joined the game after the launch of OW2. In general, OW1 players expressed disappointment witnessing their beloved game fail. Many posts mentioned friends quitting due to perceived lack of motivation and little enjoyment in OW2. On the other hand, new OW2 players expressed concerns about item scarcity and envy those with expensive skins. While players acknowledged the necessity of the F2P transition for the game's long-term evolution, the current predatory practices are driving away players in OW2. In short, players believed that the current game is unsustainable long-term because, over time, only big spenders will remain in game, but they too will leave because of OW2's \textit{``lacklustre''} rewards. 

\begin{quoting}

    ``Games nowadays are not just about (core) gameplay. People expect progression and rewards for their experience.'' (Reddit post 52)

\end{quoting}

\subsubsection{Detrimental practices affecting game quality and publisher reputation}

Many posts expressed hopelessness regarding the game's future, as many players mentioned the game publisher's delay of response to tech support tickets about the reported bugs, and no compensations (as of our data collection) were provided to players who suffered from item lost in their accounts, leaving players helpless in all the problems. In addition, many posts reported the game publisher's politician-like responses to all the problems---only solving the problems that brought up the media attention to prevent social backlash and letting the rest of the problems fall on deaf ears. 

\begin{quoting}

    ``The game feels like a politician who wears a broad smile, putting on a show with a rehearsed speech and fake enthusiasm.''    (Reddit post 251)

\end{quoting}

In addition, many posts concerned that OW2 prioritizes profit at the expense of game quality, as the current OW2 removed many OW1 game features that excited players, such as skin lore and endorsing enemy players. Game shop items suddenly feel mass-produced without background stories. As specified in the posts, the game launch had many bugs and issues from insufficient preparation, including the disappearance of OW1 items in OW2 accounts, all heroes getting locked for payments, the \textit{``extremely long''} login wait times, and being \textit{``kicked out''} halfway through a game. Many players purchased the pre-launch item bundle (i.e., Watchpoint pack) but did not receive items in their accounts nor refunds. Many posts said the monetization system was the \textit{``only functional feature''} in OW2 and the F2P model were deemed as an excuse for bad game quality because \textit{``we[players] did not pay for those.''} 

\begin{quoting}

    ``OW1 had a vibrant community and healthy gameplay, but OW2 feels like a rushed and buggy product that has been in development for barely a couple months, not 3.5-7 years.'' (Reddit post 8)

\end{quoting}

The \bmt also consequentially diminished the publisher's reputation among players. As players reported in the Reddit posts, the publisher is no longer viewed as the company that crafts high-quality games that players could enjoy for years. Instead of enhancing player motivation and in-game experiences, the publisher had diverted attention to other marketing opportunities outside the game, such as offering Twitch drops of in-game items.

\begin{quoting}

    ``It's absurd that the only way to get free skins for this year's Junkenstein event is by watching others play the game, instead of earning them through our own gameplay.'' (Reddit post 72)

\end{quoting}

\subsection{From players' perspectives, what elements contribute to a satisfying game business model transition?}
\label{subsubsec:RQ3}

\subsubsection{Improve reward and progression systems}

To answer our RQ2, we analyzed Reddit posts to further reveal players' desires to fix the issues and make the game more accessible and sustainable in the long term. For instance, we found players' expectation of an improved reward system that respects players' investments of time, money, cognitive effort. Recommendations included providing default rewards for dedicated weekly or daily players instead of expecting them to pay \$10 per season as a pseudo-subscription. Some posts suggested introducing a new progression system separate from \varname{\BP} to solve the lack of progression problem for free players. This could be an account-based level progression or character-based progression tied to players' use. Specifically for holiday events, players suggested to offer skins as rewards of completing event challenges, as the current rewards (e.g., sprays, souvenirs, or weapon charms) are not something people care about and want to earn.

\begin{quoting}

    ``If a game treats its players well and offers cool skins for my favorite characters, I'm more tempted to make a purchase.'' (Reddit post 319)

\end{quoting}

Players also expected more accessible rewards, such as being able to earn \varname{OW2 coins} through completing \varname{\BP} tiers, and having options to earn alternative rewards instead of the mythic skin. This way could mitigate players' frustration from \textit{``useless''} rewards, ease their grinding burden, and encourage them to buy as they are less afraid to waste coins. In addition, players also expect a better utility of the coins from the OW1, such as for purchasing old OW1 items instead of just for default skins. Other posts recommended not locking the new character in \varname{\BP}, as it can encourage a pay-to-win dynamic and diminishes players' excitement for the new character that they cannot play for a while.

\begin{quoting}

    ``Receiving in-game currency as a reward leads to my desire for more, as I can get my desire item by purchasing 1 additional cheap currency pack.'' (Reddit post 18)

\end{quoting}

\subsubsection{Embrace player feedback and fair business practices}

On the one hand, players are eager for the game publisher to listen to community feedback and actively respond or implement changes to decrease the players' negative perceptions. In this regard, many posts proposed strategies to assist the game publisher in gaining a deeper understanding of player preferences, such as having collaborative efforts to systematically categorize, compile, and prioritize player feedback. On the other hand, a few posts expressed players' hopelessness that their feedback will never be heard. These players advocated boycotting and stopping playing the game until improvements are made, as they believed that only a tangible decline in revenue would prompt the game publisher to consider and address player feedback.

Finally, many posts suggested to find a balance between the necessity for revenue generation and avoiding player exploitation, ensuring that players have enjoyable and satisfying gaming experiences while also sustaining the game financially for the long term. A common sentiment expressed in these posts advocated for fair pricing of in-game shop items, particularly those exclusive to a single character. Overall, players are inclined to support a game that demonstrates respect and care for its player base.  

\begin{quoting}

    ``Players are more likely to spend money on games from companies they trust. I can't support a game that feels solely driven by greed.'' (Reddit post 319)
\end{quoting}

%% file: tables/Table-II.tex
\begin{table}[!t]
\centering
\caption{Deceptive design patterns~\cite{zagal2013dark} demonstrated in OW2 game mechanics, which served as a starting point for our qualitative analysis on players' perceptions and experiences from Reddit.}
\label{tab:UI-analysis}
\resizebox{\textwidth}{!}{%
\begin{tabular}{@{}llll@{}}
\toprule
\textbf{Category}~\cite{zagal2013dark} & \textbf{Deceptive Pattern}~\cite{zagal2013dark} & \textbf{OW2 Game Mechanics*} & \textbf{Short Description} \\ \midrule
Temporal & Grinding & \cellcolor{paleblue!15}Daily/Weekly/Seasonal Challenges& \cellcolor{paleblue!15}\textcolor{paleblue}{\faChevronCircleRight}~Players have the opportunities to complete 3/11/41 repetitive tasks on a daily/weekly/ \\ 
Patterns & & \cellcolor{paleblue!15} & \cellcolor{paleblue!15}seasonal basis, such as ``earn 10 eliminations/assists without dying,'' ``win 10 games queued \\
& & \cellcolor{paleblue!15} & \cellcolor{paleblue!15}as all roles,'' and ``win 100 games in competitive play.'' \\ 
 &  & \cellcolor{palegreen!15}Battle Pass & \cellcolor{palegreen!15}\textcolor{palegreen}{\faChevronCircleRight}~Players advance in the battle pass by playing the game and completing daily and weekly \\ 
 & & \cellcolor{palegreen!15}& \cellcolor{palegreen!15}challenges, earning XP. Each battle pass level necessitates 10,000 XP. \\ 
 &  & \cellcolor{paleorange!15}New character & \cellcolor{paleorange!15}\textcolor{paleorange}{\faChevronCircleRight}~New characters are introduced every two seasons. If players fail to complete the battle \\
  & & \cellcolor{paleorange!15}& \cellcolor{paleorange!15}pass for a specific season, they must complete 7 special challenges (e.g., ``Win 35 games \\
  & & \cellcolor{paleorange!15}& \cellcolor{paleorange!15}queued as All Roles or playing Tank heroes in any game modes.'') to acquire it. \\ 
 & Infinite Treadmill & \cellcolor{paleblue!15}Daily/Weekly/Seasonal Challenges& \cellcolor{paleblue!15}\textcolor{paleblue}{\faChevronCircleRight}~The progress resets daily/weekly/every season (9 weeks~\cite{OW2seasonduration}). \\
  &  & \cellcolor{palegreen!15}Battle Pass & \cellcolor{palegreen!15}\textcolor{palegreen}{\faChevronCircleRight}~A new battle pass is introduced every season (9 weeks~\cite{OW2seasonduration}). \\ 
 & Playing by Appointment & \cellcolor{palepink!15}Double XP Weekend &  \cellcolor{palepink!15}\textcolor{palepink}{\faChevronCircleRight}~
A weekend event from March 10 to 12, 2023, celebrates and promotes the upcoming \\ 
 & &\cellcolor{palepink!15} &\cellcolor{palepink!15}PACHIMARCHI event. To participate, players must play within this time frame. \\ 
 &  & \cellcolor{palepink!15}PACHIMARCHI event & \cellcolor{palepink!15}\textcolor{palepink}{\faChevronCircleRight}~
The event spans two weeks, from March 21 to April 4, and participation requires playing \\
 & & \cellcolor{palepink!15} &  \cellcolor{palepink!15}within that timeframe. \\ \midrule
 Monetary & Anchoring Tricks & \cellcolor{palegreen!15}Battle Pass & \cellcolor{palegreen!15}\textcolor{palegreen}{\faChevronCircleRight}~A bundle of 80 items purchasable with 1000 in-game coins (equivalent to \$10 USD value). \\ 
 Patterns & &\cellcolor{palered!15}In-Game Shop & \cellcolor{palered!15}\textcolor{palered}{\faChevronCircleRight}~Game item bundles. \\ 
& Recurring Fee & \cellcolor{palegreen!15}Battle Pass & \cellcolor{palegreen!15}\textcolor{palegreen}{\faChevronCircleRight}~The \$10 USD premium encourages players to play more to get their money's worth. \\
& Artificial Scarcity & \cellcolor{palered!15}In-Game Shop & \cellcolor{palered!15}\textcolor{palered}{\faChevronCircleRight}~Game shop items reset weekly. \\ 
& Pay to Skip & \cellcolor{palegreen!15}Battle Pass & \cellcolor{palegreen!15}\textcolor{palegreen}{\faChevronCircleRight}~Players can purchase tiers to bypass the grind and obtain items from the battle pass. \\ 
& Waste Aversion & \cellcolor{palered!15}In-Game Shop & \cellcolor{palered!15}\textcolor{palered}{\faChevronCircleRight}~OW coins can be purchased in different amounts than what players can spend to buy \\
& & \cellcolor{palered!15}& \cellcolor{palered!15}in-game items. The unused coins generate a sense of money being wasted, prompting \\
& & \cellcolor{palered!15}& \cellcolor{palered!15}players to buy more coins. \\
& Premium Currency & \cellcolor{palered!15}OW coins & \cellcolor{palered!15}\textcolor{palered}{\faChevronCircleRight}~Coins in OW2 obscure the true USD spending and by-pass laws. In OW2 coins are now \\
& & \cellcolor{palered!15}& \cellcolor{palered!15}purchasable with real money, unlike in OW1 where players can earn them in gameplay. \\ \midrule
Social  & Fear of Missing Out & \cellcolor{paleblue!15}Daily/Weekly/Seasonal Challenges& \cellcolor{paleblue!15}\textcolor{paleblue}{\faChevronCircleRight}~The progress resets daily/weekly/every season (9 weeks~\cite{OW2seasonduration}). \\
Patterns &  & \cellcolor{palegreen!15}Battle Pass & \cellcolor{palegreen!15}\textcolor{palegreen}{\faChevronCircleRight}~A new battle pass is introduced every season (9 weeks~\cite{OW2seasonduration}). \\ 
  &  & \cellcolor{paleorange!15}New character & \cellcolor{paleorange!15}\textcolor{paleorange}{\faChevronCircleRight}~New characters are introduced every two seasons. If players fail to complete the battle \\
  & & \cellcolor{paleorange!15}& \cellcolor{paleorange!15}pass for a specific season, they must complete 7 special challenges (e.g., ``Win 35 games \\
  & & \cellcolor{paleorange!15}& \cellcolor{paleorange!15}queued as All Roles or playing Tank heroes in any game modes.'') to acquire it. \\ \midrule
  Psychological & Aesthetic Manipulation & \cellcolor{palegreen!15}Battle Pass & \cellcolor{palegreen!15}\textcolor{palegreen}{\faChevronCircleRight}~Level 80 rewards a mythic skin with customizable attire and unique audio effects that \\ 
  Patterns & & \cellcolor{palegreen!15}& \cellcolor{palegreen!15}other skins do not have. \\ 
  & Endowed Progress & \cellcolor{paleblue!15}Daily/Weekly/Seasonal Challenges & \cellcolor{paleblue!15}\textcolor{paleblue}{\faChevronCircleRight}~The individual challenges, like ``heal a total of 65,000 damages,'' and ``win 100 games in \\ 
 & & \cellcolor{paleblue!15} & \cellcolor{paleblue!15}competitive mode,'' along with the opportunity to complete 3 challenges daily/weekly or 41 \\
 & & \cellcolor{paleblue!15} & \cellcolor{paleblue!15}challenges every season, create compelling artificial goals that players find hard to abandon. \\ 
  & & \cellcolor{palegreen!15}Battle Pass & \cellcolor{palegreen!15}\textcolor{palegreen}{\faChevronCircleRight}~
The 80 levels of the battle pass and their corresponding required XP create a compelling \\
  & & \cellcolor{palegreen!15} & \cellcolor{palegreen!15}objective that players find hard to abandon. \\ \bottomrule
\multicolumn{4}{l}{\textit{Note}. *Patterns shared in common between OW1 and OW2 are excluded. Related game mechanics are grouped by colour.} \\
\end{tabular}%
}
\end{table}

%% file: sections/05-Discussion.tex
Our analyses of OW2 game mechanics and player perceptions identified the use of deceptive design practices during the game's \bmt and their consequential impact on players. In this section, we first compare our study with other deceptive game design studies, we then discuss the implication for future research on deceptive design and games, and ethical game design. 

\subsection{Implications for Deceptive Design and Games Research}

\subsubsection{Assessing Deceptiveness Relies on Player Perception}

Our two-phased methodology allowed us to empirically ground how we assessed deceptive game mechanics on player interactions, perceptions, and experiences. This approach enabled us to reduce the ``subjectivity'' in~\citet{zagal2013dark} and identify elements that genuinely contribute to players' feelings of being manipulated during gameplay~\cite{deterding2020against}. For example, \textit{Grinding} is a temporal design with mixed effects~\cite{zagal2013dark,karlsen2019exploited}. The repetitiveness of grinding attempts to conceal players' time investment. It could also ease immersion and engagement for players who find enjoyment in repetitive tasks. These players may also feel satisfaction in accumulating resources or curiosity about potential future changes~\cite{karlsen2019exploited,zagal2013dark}. However, in games like OW2, players confirmed that the grind experienced was tedious, boring, and compulsive rather than fun, intriguing, and voluntary. This observation led us to the conclusion that the \textit{Grinding} patterns in OW2 are deceptive design.

We adopted the~\citet{zagal2013dark}'s taxonomy as a starting point for our game mechanics analysis. However, we found that assessing the harmful effects of deceptive designs required us to understand player perceptions of deceptive practices in games. Examples include players feeling socially obliged to unlock a new character and the accidental purchases from the game's shop user interface design. On the other hand, we found that several designs that seemed deceptive in nature were ineffective on players because of confounding factors in the game. For instance, \textit{Anchoring Tricks} and \textit{Aesthetic Manipulations} appeared ineffective in influencing OW2 players behaviour caused by the perceived ``lackluster'' quality of in-game items.

Our research suggests that players' awareness and experience of deceptive design in video games influences how they detect deceptive game mechanics. This shows the importance of the recommendations from~\citet{gray2021enduser,gray2023mapping} to incorporate player perception analysis.  
As research on deceptive game design continues to expand, we encourage future research to incorporate player perceptions and experiences. This will enhance their classification and assessment of deceptive design in games. Future game researchers and designers could also uncover insights into alternative ``bright patterns'' from player expectations and feedback~\cite{sandhaus2023promoting}. These patterns foster player motivation and game playfulness without harmful outcomes.

\subsubsection{Comparing OW2 to Other Deceptive Game Design Taxonomies}

Some of our findings echoed previous research on player perceptions of deceptive game design (e.g.,~\cite{fitton2019F2P,karlsen2019exploited}). For example, we found that players feel pressured by \textit{Grinding} to spend more time in game, \textit{Artificial Scarcity} caused players to spend more money in the game shop, and premium fees that restricted game progression drove players to buy the \BP (see~\autoref{tab:RQ3} in Appendix). 

In addition, we identified many deceptive practices that fall beyond the scope of~\citet{zagal2013dark}'s classification but align with the taxonomies in more recent research~\cite{petrovskaya2021predatory,fitton2019F2P,roffarello2023defining,king20233d}. For example, the OW's act of freely giving away shop items through Twitch drops is better aligned with the \textit{Separate Re-Release of Product as Free, Cheaper, or Easier to Get} pattern found by~\citet{petrovskaya2021predatory}. Similarly, the vague information about the utility of OW1 coins in OW2 matches the \textit{Lack of Information About Condition of Product} pattern in~\citet{petrovskaya2021predatory}. Furthermore, we identified certain patterns related to, but do not squarely fit within, the established classifications by~\citet{petrovskaya2021predatory} and~\citet{king20233d}. For instance, the OW2 reward system might also align with the \textit{Real Money Spend Expected} pattern in~\citet{king20233d} because players believed that high-quality rewards cannot be realistically obtained without spending money. While the game remains playable in a technical sense, the lack of progress without premium \varname{\BP}, significantly diminished player satisfaction. Thus, OW2's reward system might partially deploy the deceptive pattern, \textit{Game Unplayable Without Spending Money} in~\citet{petrovskaya2021predatory}, by making the experience feel meaningfully unplayable unless players invest real money.

In addition, our analysis on player perceptions revealed problematic practices that had previously been overlooked in the literature on deceptive game design. Examples include price skimming and the publisher's evasive response to player feedback. 

These results suggests the need for a more comprehensive classification that specifically addresses deceptive design in video games. While literature has begun to synthesize deceptive designs in existing research, they have been discussed broadly together with other digital interfaces~\cite{gray2023mapping,gray2018dark,roffarello2023defining} or predominately focused on mobile game patterns that relied on non-game features in mobile apps (e.g., push notification-based ads~\cite{roffarello2023defining}). These studies did not capture deceptive game design mechanism emerging from game development, and player interactions, perceptions, and experiences that are unique to PC-based gameplay contexts. Thus, there is a need to develop a comprehensive classification of deceptive design in PC-based gaming environments. Further, deceptive design in mobile games, console-based games and VR games may need more granular classification to capture deceptive design practices arising from game features unique to the platforms. 
Research could expand beyond deceptive game interface design elements to include manipulation that result from players' interaction sequence and the game publisher's problematic business practices. 

\subsection{Implications for Ethical Game Design and Satisfying Game Model Transition}

Previous research suggests that transitioning to a F2P business model benefits the game and the publisher (e.g.,~\cite{rizani2020analysis,newham2022consequences}), but our findings from the players' perspective show that the use of deceptive and unethical practices can harm the player experience and the game over time. Our observed issues, player perceptions, and suggested remedies aim to guide future game development to create a more enjoyable and satisfying game \bmt. In this section, we draw implications from our findings for designers, researchers, and publishers in game industry and for future games that considering a F2P transition.

\subsubsection{Ensuring Fundamental Player Motivating Components During Game Model Transition}
Our analysis identified significant issues during the transition that have resulted in the removal of fundamental game components~\cite{hogberg2019gameful,yee2006motivations,toups2016collecting} that motivated and engaged players in OW1. For instance, the lootboxes in OW1 fulfilled players' feelings of progression from receiving level-up rewards, and the ability to unlock new items for their frequently played character. However, the deceptive game mechanics in OW2 made these progressions inaccessible without payments, leading to frustration among players and challenging the traditional notions of game achievement and success~\cite{hogberg2019gameful, newham2022consequences, luton2013free}. In addition, the \varname{Daily/Weekly/Seasonal Challenges} and the \varname{\BP} system obligate users to grind in a certain way, instead of guiding them to explore it voluntarily and pleasurably like in OW1. This diminished the overall playfulness of game experience~\cite{hogberg2019gameful}. The use of false impressions through \textit{``fake''} bundles and discounts were perceived more than inconvenience to drive players' desire of in-game items~\cite{hamari2010game, evans2016economics} but as predatory~\cite{petrovskaya2021predatory, king2018predatory} and violations of Consumer Protection laws\footnote{Australian Competition and Consumer Commission. \url{https://www.accc.gov.au/consumers/buying-products-and-services/consumer-rights-and-guarantees}}\textsuperscript{,}\footnote{Canadian False or Misleading Representations and Deceptive Marketing Practices.\url{https://ised-isde.canada.ca/site/competition-bureau-canada/en/deceptive-marketing-practices/types-deceptive-marketing-practices/false-or-misleading-representations-and-deceptive-marketing-practices}}~\cite{king2019unfair}. To ensure the autotelic aspect of gameplay for players, our findings suggest that while new game mechanics are necessary, future games should maintain the fundamental elements that motivate and engage players during their business model transitions~\cite{hogberg2019gameful,yee2006motivations,toups2016collecting}. Designers should give various rewards to cater to different players' interests as F2P expands the game's playerbase. Game designers and publishers must also consider ethics and avoid violating consumer protection laws or exploiting players' vulnerabilities. Misleading and dishonest promotional statements degrade player trust over time. Finally, we caution publishers against using monetization tactics because predatory implementations can diminish the allure of ``free'' gameplay and negatively impact player satisfaction and engagement~\cite{petrovskaya2021predatory,king2018predatory,zagal2013dark}. Game designers should meticulously evaluate each player-motivating pattern within the broader game design, as unintended and potentially conflicting consequences can arise from their interactions. 

\subsubsection{Balancing Player Investments and Fairness of Rewards}
Our Reddit analysis revealed frustration, disappointment, and disengagement among both new OW2 players and OW1 veterans. These reactions were rooted in the inbalance between players' investments (i.e., time, money, cognitive effort) and the rewards they received in return. The literature suggests that the fairness of games is influenced by gameplay balance and appropriate matchmaking~\cite{sanchez2022welfare}, equal resource accessibility~\cite{freeman2022pay}, and the integration of exploitative in-game purchase features~\cite{king2019unfair,petrovskaya2021predatory}. Our results indicate that as a crucial part of game fairness, players want a balance between their investments in the game and the quality, utility, and value of their compensation rewards. Players invested more in the OW2 due to time sinks, money sinks, and social and psychological tactics, but the rewards for these investments did not meet their expectations (e.g., useless, or meager rewards). In addition, players also expressed dissatisfaction when OW2 compromised their ability to obtain items that were freely available in OW1. Therefore, to promote game fairness, we suggest that game publishers should appropriately compensate player investments. A fair reward system that recognizes players' loyalty and contributions can make them feel valued and respected, leading to increased game reputation, player retention, and word-of-mouth recommendations~\cite{livingston2011impact,zhao2009factors,liao2013can}. 

\subsubsection{Fostering Player Trust and Engagement through Transparent Communication}

As discussed in~\autoref{subsubsec:RQ3}, players expected clear and open communication about changes in the games, especially when transitioning from a B2P to a F2P model. While players appreciated the developers' dedication to an enjoyable experience, the publisher's \textit{``politician-like''} response to player feedback exaggerated their perception of an unethical, unfair, and problematic game system. Research suggests that transparent communication fosters positive view, trust, and understanding between game developers and the player community~\cite{freeman2022pay}, and ensures long-term player support~\cite{vitell2003consumer, vitell1992consumer, alha2014free}. Our findings further highlight that transparent and open communication are crucial elements for maintaining a satisfactory player experience during a game's \bmt. This emphasizes the urgency for gaming companies to have a team of user researchers and community managers to actively learn the problems raised by players and develop solutions. These active actions can ensure developers are aware of player sentiment and can proactively address concerns throughout the \bmt. At the same time, it is an opportunity for developers to openly discuss the reasons behind the transition and changes to the game. By actively learning from players' feedback and community discussions, an appreciative reward systems can also be built, adjusted, and refined to meet player expectations. \discussion{Ultimately, the active learning and} transparent communication will help to foster an environment where players feel valued and encouraged to continue their journey. 

\subsubsection{Shaping Industry Ethics and A Broader Discussion on Player Rights}

Our findings revealed the wider impact of the F2P business model in relation to deceptive game on the gaming community's view of the industry. OW2's deceptive practices have sparked active debates and ignited ethical concerns about the gaming industry as a whole. The reduced player engagement and investment according to Blizzard's 2023 second-quarter financial report signifies that practices that benefit the publisher at the expense of players are unsustainable in the long term~\cite{Blizzrd2023finreport}. The ``overwhelmingly negative'' rating after OW2's launch on Steam reinforced the unsustainable practices\footnote{As of August 18, 2023. Overwatch 2 on Steam has received 144,730 reviews from players, with 131,108 being negative (91\%), resulting in a ``overwhelmingly negative'' rating towards the game overall and making it the top 1 worst game on Steam. See: \url{https://store.steampowered.com/app/2357570/Overwatch_2/} and \url{https://steam250.com/bottom100}}. This ripple effect demonstrates how individual game design decisions can have far-reaching consequences and shaping public perception and trust in the industry as a whole. It also reflects the growing concern for ethical design among players and contributes to discussions on player rights in the gaming industry.

\subsection{Limitations}

The limitations of our study are as follows. First, our researchers included an OW player with six years experience. As a result, their perspective shapes and constructs the findings. While we see it as a strength in identifying deceptive design and its impact on players, it may also introduce a limitation in terms of potential bias. Second, we analyzed the game mechanics and Reddit posts during Seasons 1\&2 of Overwatch 2 (Oct 2022 to Dec 2022). While these data allowed us to study players' immediate reactions after the \bmt and the implementation of deceptive game mechanics, we acknowledge that the game could evolve and player perceptions could change. Therefore, we encourage future research to explore players opinions when the \bmt is no longer new or news. Third, we note that our player data only came from Reddit. While Reddit offers a valuable platform for player discussions, its user demographics are skewed (i.e., as of March 2024, 75\% US, 63.6\% male, 44\% 18-29 years old)~\cite{statista2024reddit}. We tackled this problem by having a team with diverse genders and age ranges, which resulted in insights and angles of data interpretation from a diverse perspective. However, we acknowledge that the demographic of post creators may still limit the generalizability of our findings to the entire player population. Fourth, the Reddit posts we analyzed seem to lean towards more negative opinions because we limited our scope to posts about deceptive game mechanics that had received high Reddit scores. Although they reflect a major trend in the Reddit communities we studied, literature suggests that players’ perceptions of deceptive game design and a game's \bmt can vary~\cite{frommel2022daily,zagal2013dark,rizani2020analysis}. Therefore, to have a more comprehensive view of the game's \bmt and the impacts of deceptive game design on players, we encourage future research to include a sample from a broader demographic range of players and diverse online communities with a balanced negative, neutral, and positive data. This approach helps study positive and negative aspects of game design, promoting fair and ethical practices while avoiding those players find unfair. Fifth, our Reddit data was gathered using keyword search. While we carefully considered keywords from deceptive design literature (e.g.,~\cite{zagal2013dark,mathur2021makes}), it is possible that we have excluded posts where players are unaware of being manipulated and those that do not care. Future research could carefully consider methodology to capture the impacts of more insidious deceptive design. Sixth, our implications were drawn from our research on OW2's \bmt, we recommend future research to verify our findings with other game player communities or alternative data sources. Finally, we also acknowledge that there might be underlying deceptive techniques that neither current researchers nor players discovered in their gameplay experiences. Currently, there is not a comprehensive list of deceptive design used in games. While deceptive design practices may continue to emerge, this study offers a valuable contribution to the ongoing discourse. We hope it drives further research and contributes to a more exhaustive understanding and categorization of deceptive design in the future.

%% file: sections/06-Conclusion.tex
Using OW game series as an example, our study sheds light players' perspective on the role of deceptive design practices in the game's \bmt and its impact on players. The analysis of game mechanics and player experiences revealed 12 deceptive patterns in nine game mechanics caused by the transition, factors causing negative gameplay experiences, and problematic publisher practices. Our findings suggest the need for game designers and publishers to balance player investments and fairness of rewards, maintain fundamental player motivation components, and ensure transparent communication. Furthermore, the study emphasizes the importance of player perception in the classification of deceptive game design and future research. We hope our findings will serve as a valuable resource for designers, researchers, and publishers in the game industry to promote fair and transparent game design practices.

%% file: sections/98-Appendix.tex
\newpage
\section{Appendix --- Framework that Informed Our Classification of OW2 Deceptive Design} 
\input{tables/Table-III}

\newpage
\section{Appendix  --- Example of Our Game Mechanics Analysis Process}
\label{sec:app-sample-coding}

\begin{figure}[!ht]
  \includegraphics[width=\textwidth]{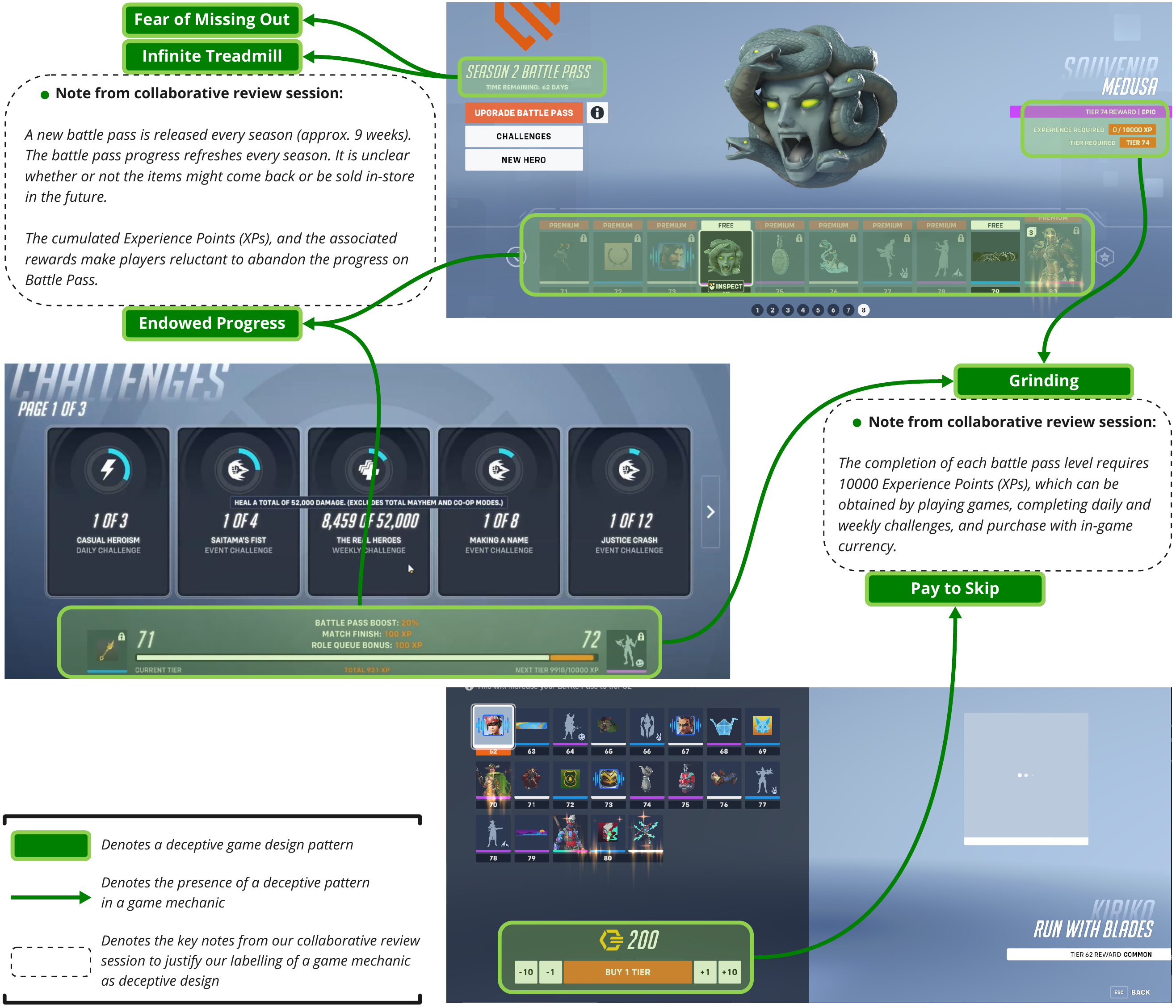}
  \vspace{-2mm}
  \caption{\UImethod{Example of our game mechanics labelling process on screenshots of OW2 \BP. Only key screenshots are included in this example. From top to bottom: (1) core \BP interface demonstrating rewards, XP requirements, and duration; (2) XP earning breakdown illustrating XP earning sources; (3) \BP level purchase interface with prices in OW coins that are purchasable with real money. Green labels denotes specific deceptive game design patterns from~\citet{zagal2013dark}. Green arrows and highlighted areas denote the presence of these deceptive game design in the game mechanics. In the deductive labelling process, we wrote down the ``Notes'' from our researchers during the collaborative review session to support labelling decisions.}}
  \Description{Example of our game mechanics coding process on a screenshot of OW2 Battle Pass.}
  \label{fig:UIexample-part1}
\end{figure}

\begin{figure}[!ht]
  \includegraphics[width=\textwidth]{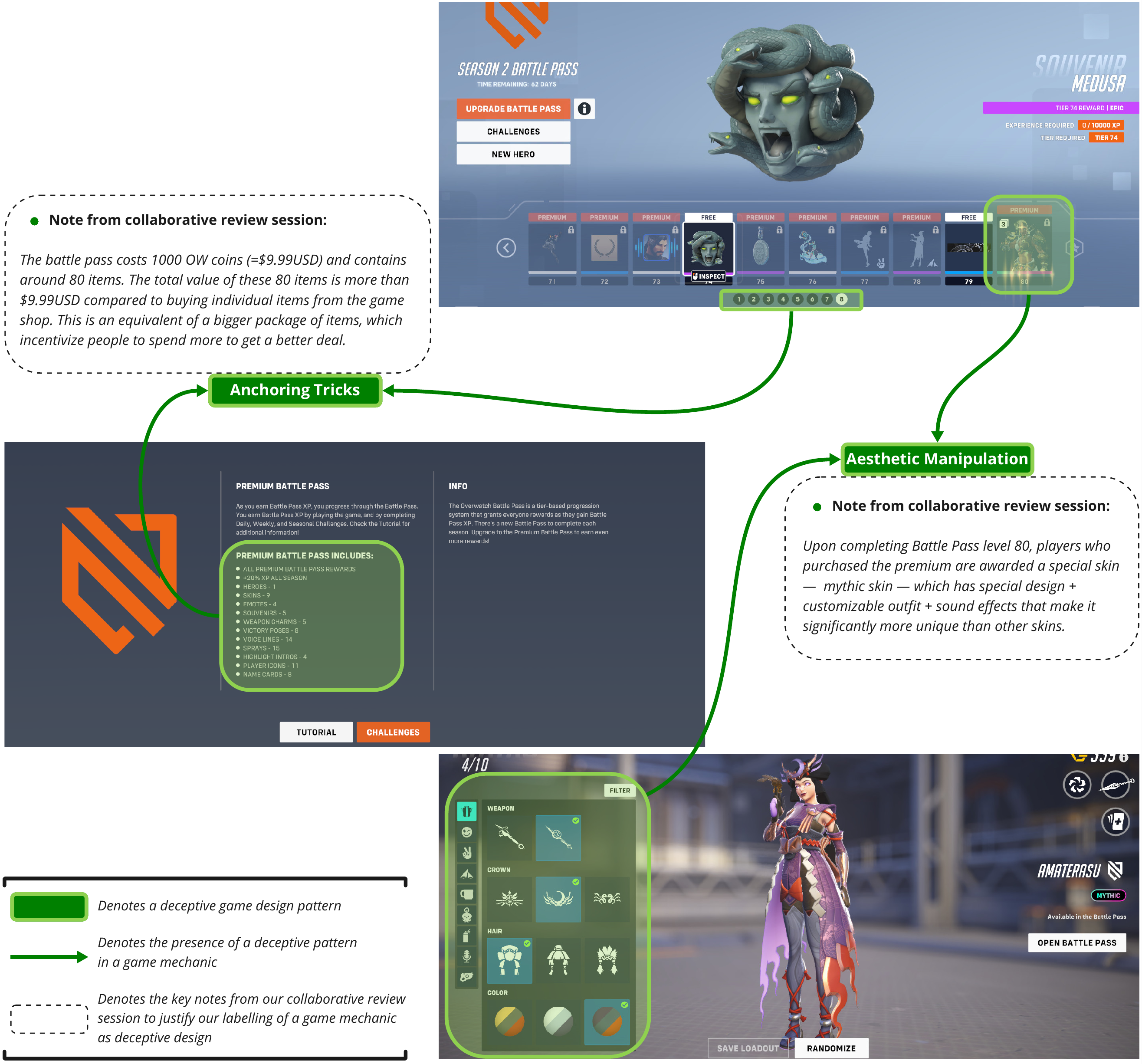}
  \vspace{-2mm}
  \caption{\UImethod{Example of our game mechanics labelling process on screenshots of OW2 \BP. Only key screenshots are included in this example. From top to bottom: (1) core \BP interface demonstrating rewards, XP requirements, and duration; (2) \BP information page detailing rewards and premium benefits; (3) mythic skin interface highlighting it's unique and customizable design. Green labels denotes specific deceptive game design patterns from~\citet{zagal2013dark}. Green arrows and highlighted areas denote the presence of these deceptive game design in the game mechanics. In the deductive labelling process, we wrote down the ``Notes'' from our researchers during the collaborative review session to support labelling decisions.}}
  \Description{Example of our game mechanics coding process on a screenshot of OW2 Battle Pass.}
  \label{fig:UIexample-part2}
\end{figure}

Here, we present an example of our game mechanics analysis on OW2 \BP in~\autoref{fig:UIexample-part1} and~\autoref{fig:UIexample-part2}. In this process, we started with the core \BP interface (top screenshot) and expanded on our discussion to other related interfaces that players could encounter during their interaction.

\clearpage
\newpage
\section{Appendix --- Reddit Search Keywords}
\label{sec:app-reddit-keywords}

\begin{itemize}
    \item 42 deceptive design keywords: [`deceptive', `deceive', `manipulative', `coercive', `malicious', `misleading', `obnoxious', `seductive', `steering', `trickery', `attack', `monetization', `money', `monetizing', `confuse', `exploit', `manipulate', `mislead', `subvert', `intent', `preference', `trick', `autonomy', `undermine', `consent', `abuse', `harm','deceiving', `manipulating', `subverting', `tricking', `abusing', `harming', `undermining', `attacking', `confusing', `scummy', `scam', `scammy', `respect', `respectful', `respective'];
    \item 21 game mechanics keywords: [`battle', `battlepass', `pass', `daily', `dailies', `challenge', `challenges', `weekly', `weeklies', `double', `event', `seasonal', `season', `coins', `cosmetics', `skin', `hero', `kiriko', `lifeweaver', `flowerman', `ramattra'].
\end{itemize}

\newpage
\section{Appendix --- Thematic Analysis Codebook of Reddit Posts}
\input{tables/Table-IV}



\newpage
\section{Appendix --- OW2 In-Game Items Glossary}
\input{tables/Table-glossary}

%% file: tables/Table-III.tex
\begin{table}[!th]
\centering
\caption{This table presents the deceptive design patterns in the design of video games by~\citet{zagal2013dark}. We used these patterns in the identification of deceptive designs in OW2.}
\label{tab:game-deceptive-design}
\resizebox{0.65\textwidth}{!}{%
\begin{adjustbox}{angle=90}  
\begin{tabular}{@{}lll@{}}
\toprule
\textbf{Category} & \textbf{Pattern} & \textbf{Description} \\ \midrule
Temporal Patterns & Grinding & Patterns that requires players to perform repetitive tasks. \\
 & Daily Rewards & Patterns that encourage players to revisit the game daily. \\ 
& Playing by Appointment &  Patterns that forces players to adhere to the game's schedule. \\
&Infinite Treadmill & Patterns that continually expand the game to never allow players to complete. \\
& Can't Pause or Save & Patterns that force players to continue playing before they can save progress.\\
& Wait To Play & Patterns that impose arbitrary wait times on players through in-game timers. \\
Monetary Patterns & Pay to Skip & Patterns that allows players players to pay to bypass undesired obstacles.\\
& Premium Currency & Patterns that disguises item real prices by introducing exchange rates.\\
& Pay to Win & Patterns that grant advantages to players who spend money.\\
& Artificial Scarcity & Patterns that use limited-time offers to manipulate player decisions.\\
& Accidental Purchases & Patterns that use visual design or default options to deceive players into making unintended purchases. \\
& Recurring Fee & Patterns that encourages players to maximize playtime to justify their spending. \\
& Gambling / Loot Boxes & Patterns that encourages players to spend money on chance-based rewards.\\
& Power Creep & Patterns that diminishing the value of purchased items over time to drive new purchases.\\
& Pay Wall & Patterns that make the game unplayable without payment.\\
& Waste Aversion & Patterns set small differences between in-game currency and item costs, prompting additional currency purchases.\\
& Anchoring Tricks & Patterns that place inexpensive items next to expensive ones to create the illusion of affordability.\\
Psychological Patterns & Invested / Endowed Value & Patterns that make it hard for players to discard investments in time and money.\\
& Badges / Endowed Progress & Patterns that discourage players from abandoning partially completed goals.\\
& Complete the Collection & Patterns that capitalize on players' strong desire to achieve in-game collections and achievements.\\
& Illusion of Control & Patterns that deceive players about their skill level to encourage more gameplay.\\
& Variable Rewards & Patterns that offer unpredictable or random rewards to enhance game addiction. \\
& Aesthetic Manipulations & Patterns that subtly influence players' aesthetic preferences through text, graphics, or sounds.\\
& Optimism and Frequency Biases & Patterns that exploit cognitive biases and clustering illusions to overestimate event frequencies (e.g., winning streaks). \\ 
Social Dark Patterns & Social Pyramid Scheme & Patterns that reward players for inviting friends.\\
& Social Obligation / Guilds & Patterns that create a sense of obligation to avoid letting down friends in the game.\\
& Friend Spam / Impersonation & Patterns that send spam to players' contact lists or social media accounts.\\
& Reciprocity & Patterns that instill a sense of obligation to reciprocate by donating resources to other players.\\
& Encourages Anti-Social Behavior & Patterns that incentivize players to engage in dishonest or harmful actions to gain an advantage.\\
& Fear of Missing Out & Patterns that induce anxiety that players will miss exciting or important events if they stop playing.\\
& Competition & Patterns that make players against each other in competition.\\ \bottomrule
\end{tabular}%
\end{adjustbox}
}
\end{table}

%% file: tables/Table-IV.tex
\begin{table}[!th]
\centering
\caption{Codebook of players experiences and perceptions toward the candidate deceptive design and OW's business model transition, found in the Overwatch community discussions.}
\label{tab:RQ3}
\resizebox{\textwidth}{!}{%
\begin{adjustbox}{angle=90}  
\begin{tabular}{@{}llll@{}}
\toprule 
\textbf{Theme} & \textbf{Code} & & \textbf{Player Statement (paraphrased*)}\\ \midrule
Time& Insane amount of grind & &Years of grind for a single skin seem insurmountable for casual players like me. \\
Sinking & New character locked behind grind & & Locking the new character in the battle pass assists grinds, nobody knows if they would \\
Experiences & & & be able to unlock her in the future\\
& Forced to play & & I feel being forced to play instead of enjoying a quality game. I have purchased content\\
& & & but still have to grind to receive rewards slowly.\\
& Huge gap between earnings and costs & & Earning 60 coins per week seems pathetic considering a legendary skin costs 1900 coins.\\
& No attainable OW coins & & I cannot earn enough coins for the next battle pass like in most games.\\
& Hard to complete & & Challenges force you to play in 4 different game modes is annoying, un-fun, and difficult.\\
Money& No progression outside of premium & & I cannot get progression, good rewards, accomplishments, recognition without premium. \\
Sinking & Removed excitement towards events & & I'm no longer excited for yearly event or holiday skins as nothing is earnable for playing.\\
Experiences & Unfair Item Pricing & & Cosmetics are overpriced.\\
& Disrespectful to my effort & & Charging or making players grind for old unlockable items is ridiculous, while giving\\
& & &  away a shop bundle for free through Twitch drops is disrespectful.\\
& FOMO sales & & I'm pressured to buy now or lose only lifetime chance for preferred skin that may never. \\
& &  & appear in shop again \\
& Accidental purchase & & The swapped placement of Ok and Cancel led to my accidental purchase of a \$20 USD skin.\\
Social & social obligation & & My team needs me to unlock a character to combat the enemy team. \\
Consequences& Players are underappreciated & & The minimal rewards from the battle pass and challenges create an unrewarding and\\
& & & disrespectful gaming experience.\\
Psychological& Lackluster Items& & The rewards are lackluster, skin designs are underwhelming, and there is nothing worth\\
Impacts& & & grinding for.\\
& Item-value limited by the game genre & & Grinding the battle pass for the mythic skin for a character I don't play is useless.\\
& No realistic way to complete collection & & The game expects me to grind 17,000 weeks or spend \$10,226 USD to get all OW2 items.\\
Predatory& Price skimming & & Intentionally overpriced contents create a false sense of value for players. A cheaper\\
Marketing& & &  system will be introduced to feign concern for and acknowledge community feedback.\\
Strategies & Fake bundles \& discounts & & Items put on sale without being sold at full price to begin with. \\
& Being vague about OW1 coins & & Intentionally vague about OW1 currency to encourage spending before OW2,\\
& & &  incentivizing real-money purchases on OW2 contents.\\
& Limited utility of OW1 coins & &OW1 coins are now useless, limited to default legendary skins only, and can't be used to\\
& & & purchase older skins that were available for years.\\
Disappointment& Better gameplay in OW2 & & I was really impressed with the core gameplay of OW2 and I enjoyed new graphics and\\
Surpassed& & &  game modes.\\
Enjoyment& Lost the sense of rewarding & & Everything in the game feels empty and pointless without spending money.\\
& Lost interest in items & & I can feel myself just completely disconnecting with caring about cosmetics.\\
& Free players have no incentives to play & & I don't want to burst the free-to-play bubble, but don't expect to get cool items without\\
& & & spending some cash\\
& Unsustainable in the long-term & & The current system seriously injures the game's growth and lifespan. It only caters to\\
& & & whales who spend big but will quickly move on to the next trendy game. \\
Detrimental & Evasive response to player feedback & & The game publisher acts like they listen to their players but don't actually care.\\
Business & Value profit more than gameplay & & Giving out (old) skins in Twitch drops instead of in gameplay says a lot about their values\\
Practices& & &  and focus.\\
\bottomrule
\multicolumn{4}{l}{\begin{tabular}[c]{@{}l@{}}\textit{Note}. *For ethical considerations, we paraphrased all direct quotes in such a way that the original post and the creator are not easily traceable. \\See~\autoref{subsec:ethical-considerations}.\\  \end{tabular}} \\
\end{tabular}
\end{adjustbox}
}
\end{table}

%% file: tables/Table-glossary.tex
\begin{table}[!th]
\centering
\caption{OW2 In-Game Obtainable Items Glossary }
\label{tab:glossary}
\resizebox{\textwidth}{!}{%
\begin{tabular}{@{}lcl@{}}
\toprule
\textbf{Item*} & \textbf{Example} & \textbf{Description} \\ \midrule
Player Icons& \includegraphics[width=0.09\textwidth]{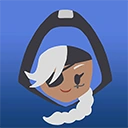}& Player icons are avatars which are displayed to a player's friends and group. \\ 
Name Cards & \includegraphics[width=0.3\textwidth]{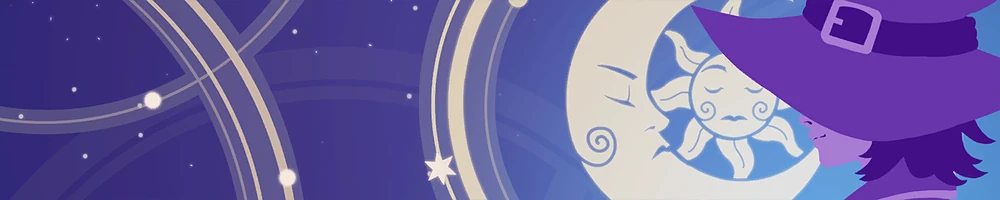} & Name cards are on the background of a player's profile. \\ 
Skins& \includegraphics[width=0.09\textwidth]{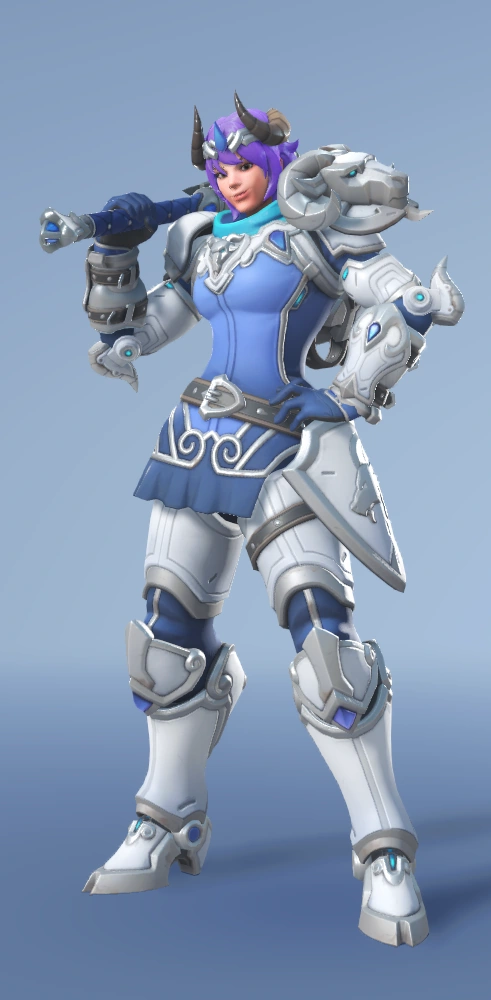}& Skins modify the appearance and voice clips of heroes.\\ 
Emotes &\includegraphics[width=0.09\textwidth]{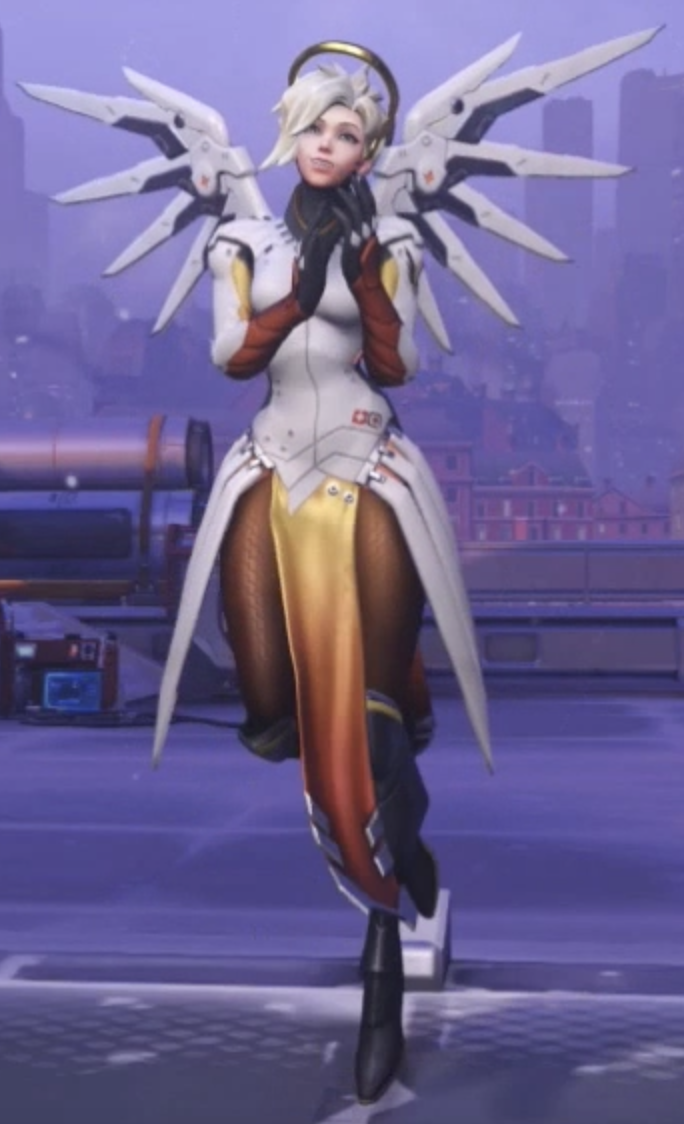}& Emotes make the hero strike a pose or perform an action while in-game.\\ 
Souvenirs& \includegraphics[width=0.16\textwidth]{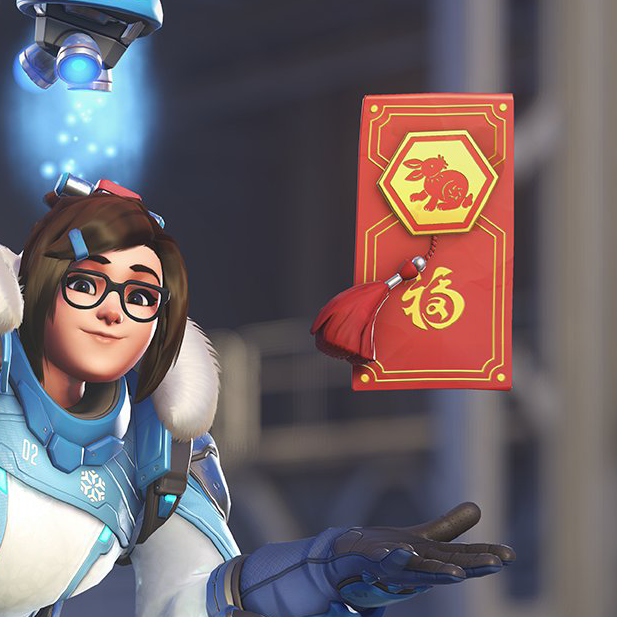} & Souvenirs are objects displayed with emotes.\\ 
Weapon Charms& \includegraphics[width=0.09\textwidth]{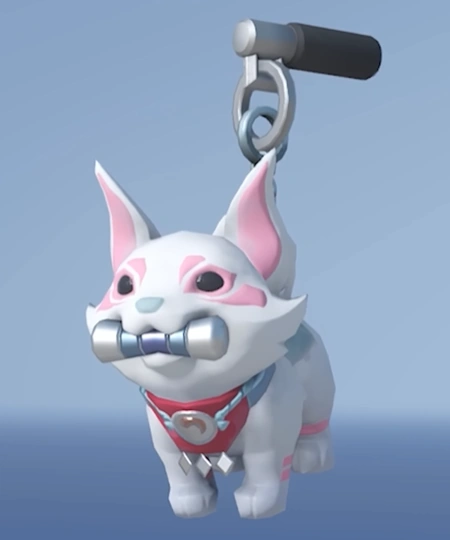}& Weapon charms can be hung on a hero's weapon.\\ 
Victory Poses& \includegraphics[width=0.17\textwidth]{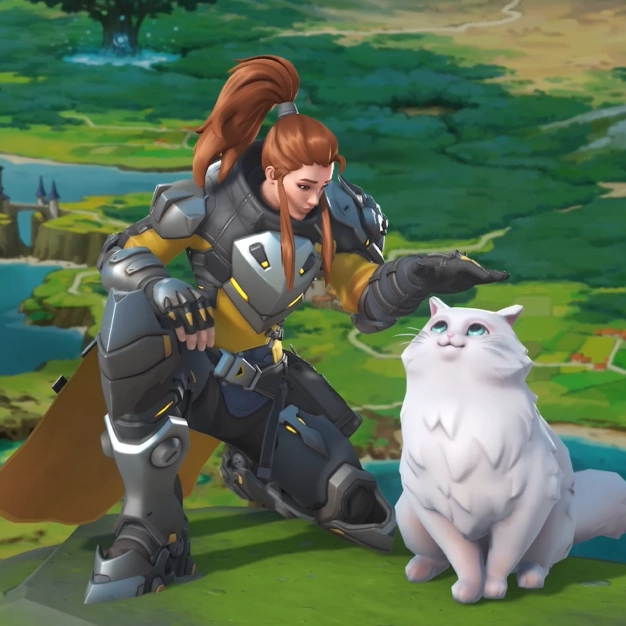} & Victory poses modify how a hero stands on the victory screen.\\ 
Voice Lines& \includegraphics[width=0.09\textwidth]{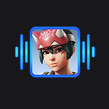} &  Voice lines make other players around the hero hear them say a phrase.\\ 
Sprays& \includegraphics[width=0.09\textwidth]{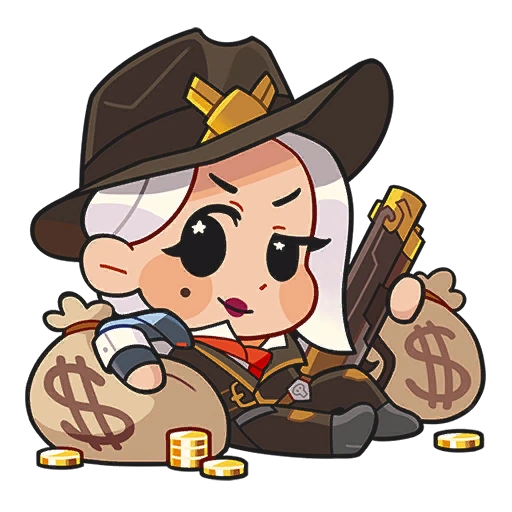}& Sprays can be put on surfaces in game by pressing the bound key or button.\\ 
Highlight Intros& \includegraphics[width=0.18\textwidth]{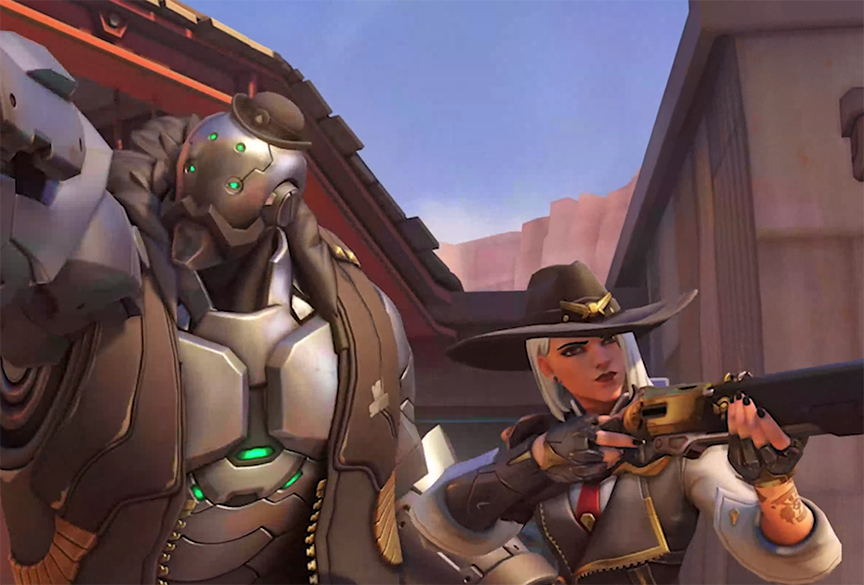}& Highlight intros customize the animation that is shown when the game ends.\\ %
\bottomrule
\multicolumn{3}{l}{\begin{tabular}[c]{@{}l@{}}\textit{Note}. Items such as \textit{player title} and \textit{golden weapon} that cannot be obtained from the \BP and the Game Shop are omitted.\\
*Item information is based on Overwatch Wiki.~\url{https://overwatch.fandom.com/wiki/Cosmetics}\end{tabular}} \\
\end{tabular}%
}
\end{table}